\documentstyle[preprint,aps,epsf]{revtex}
\tighten
\begin{document}
\newcommand{\rb}[1]{\raisebox{-1ex}[-1ex]{#1}}
\newcommand{\sst}[1]{\scriptscriptstyle{#1}}
\newcommand{\dfrac}[2]{\mbox{$\displaystyle\frac{#1}{#2}$}}

\draft

\title{
Extraction of electromagnetic neutron form factors through inclusive
and exclusive polarized electron scattering on polarized $\bbox{^3}$He target
}

\author{
J. Golak$^\dagger$, G. Ziemer, H. Kamada\footnote{present address:
Institut f\"ur Strahlen- und Kernphysik der Universit\"at  Bonn
Nussallee 14-16, D53115 Bonn, Germany}, H. Wita\l{}a$^\dagger$, W.Gl\"ockle
}
\address{Institut f\"ur Theoretische Physik II,
         Ruhr-Universit\"at Bochum, D-44780 Bochum, Germany}
\address{$^\dagger$Institute of Physics, Jagellonian University,
                    PL-30059 Cracow, Poland}

\date{\today}
\maketitle

\begin{abstract}
The inclusive and exclusive processes 
$\overrightarrow{^3{\rm He}}(\vec{e},e')$ 
and 
$\overrightarrow{^3{\rm He}}(\vec{e},e'n)$ 
have been theoretically analyzed and values for the magnetic and electric 
neutron form factors have been extracted. In both cases the form factor 
values agree well with the ones extracted from processes 
on the deuteron.  Our results are based on Faddeev solutions, modern NN forces
and partially on the incorporation of mesonic exchange currents.
\end{abstract}

\narrowtext

\section{Introduction}
\label{secIN}

Besides the deuteron, polarized $^3$He appears to be a useful target
to extract information on the electromagnetic
neutron form factors. This proposal goes back to~\cite{Blankleider} 
and has been emphasized again by~\cite{Friar}. It is based on the fact
that the principal S-state dominates the $^3$He wave function by more than
90~\% and in this state the polarization is carried solely by the
neutron. Needless to say that the knowledge of the electromagnetic form
factors of the neutron is basic to get insight into the distribution 
of charge and magnetization inside the neutron.
We refer to~\cite{Jager,Rohe,Passchier,Herberg,Anklin94,Anklin98} 
and~\cite{star2} for experimental and theoretical 
work on that topic extracting information on the neutron form factors. 
In this article we would like to analyze two experiments~\cite{doublestar,triplestar}
 on the processes
$\overrightarrow{^3{\rm He}}(\vec{e},e')$ 
and 
$\overrightarrow{^3{\rm He}}(\vec{e},e'n)$ 
with the aim to extract the magnetic and
electric neutron form factors at certain $Q^2$ values. This theoretical
analysis will be based on Faddeev solutions for the 3N continuum 
and the 3N bound state belonging to the same 3N Hamiltonian. 
We shall also use realistic NN forces. For the inclusive process 
we carried through  an analysis before~\cite{ref3}, but now it refers to 
a new more accurate experiment~\cite{doublestar}
and also the theory will be improved 
by including mesonic exchange
currents. The analysis~\cite{Ziemer} of the exclusive experiment by consistent 
Faddeev solutions for the 3N continuum and $^3$He has not been 
done before to the best of our knowledge.

In section~\ref{secII} we shall investigate the inclusive 
process and in section~\ref{secIII} the exclusive one. 
We close with an outlook in section~\ref{secIV}.

\section{Extraction of the magnetic form factor of the neutron}
\label{secII}
 
In this section we shall analyze a measurement of 
$\overrightarrow{^3{\rm He}}(\vec{e},e')$ 
carried 
through at Jlab~\cite{doublestar}.
We refer to~\cite{ref3} for the detailed theoretical formalism and restrict
ourselves to describe only its extensions. That article will henceforth
be cited as I and equations thereof by (I.*).
In I we used only a single nucleon current operator. Now we add two-body
exchange current operators. The central Faddeev-like equation given in Eq.~(I.28)
is derived under the assumption
that the operator $C$ has the form of (I.29). In I this simply meant 
that $C$ is a sum of three single particle operators. However, what really 
enters the derivation of Eq.~(I.28) is, that the operator $C$ can be decomposed 
into three parts such that $C^{(i)}$ is
symmetrical under exchange of particles $j \ne k$ with $j \ne i \ne k$.
The operator $C$ has the physical meaning of a component of the current
operator.
Thus we can simply add two-body currents, which naturally decompose 
in a 3N system into three parts and therefore $C^{(1)}$ in Eq.~(I.28) will be
now a sum of two terms:
\begin{equation}
C^{(1)} =  C^{(1)}_{\rm sing} + C^{(23)}_{\rm exch}
\label{eqci}
\end{equation}
The first term is the single-nucleon current used in I and $C^{(23)}_{\rm exch}$ 
is the corresponding component of a two-body current acting on particles 2 and 3. 
As a consequence there will
occur now an additional driving term in Eq.~(I.28) of the form 
$(1 + t G_0) C^{(23)}_{\rm exch}  \vert \Psi_{^3He} m > $.

The following steps in I concern the partial wave representation. Since
the spherical components $ C^{(23)}_{\pm 1} $ of the two-body current 
operator are tensor operators and behave like the single
nucleon components used in I, the conditions~(I.36-37) and as a consequence
(I.38) remain valid. The symmetry properties~(I.41-44)
based on the partial wave decomposed forms remain also valid for the
additional  two-body currents.
This follows from their explicit forms as given in~\cite{Kotlyar}. 
Then the following expressions
leading to the final forms of the four response functions~(I.52-55)
remain valid.

For the two-body currents we follow the Riska prescription~\cite{Riska}, 
which via the continuity equation relates NN forces and exchange currents
in a model independent manner, as it is often referred to. We choose the
AV18 NN force model~\cite{AV18} and restrict ourselves to the dominant $\pi$-
and $\rho$-like parts. 
We refer to~\cite{ref6} for more details
and to~\cite{Kotlyar} for the partial wave expansion of the two-body currents.
In case of Bonn~B~\cite{BONNB},
which we also use as another NN force model, we choose standard 
$\pi$- and $\rho$-meson exchange currents augmented by the strong form factors
used in Bonn~B. For the proton electromagnetic form factors we took 
the H\"ohler~\cite{Hoehler} parametrization, which at $Q^2$= 0.1 
and 0.2 ${\rm GeV}^2/c^2$ agrees perfectly with the data.

To theoretically analyze the data from~\cite{doublestar} the experimental conditions 
have to be taken into account. The incoming electron beam energy 
was $E$= 778 MeV. 
The central electron scattering angles for $Q^2$= 0.1 and 0.2 ${\rm GeV}^2/c^2$
were $\theta_e$= 24.44~$^\circ$ and 35.5~$^\circ$, respectively. 
The spread in the electron angles were $ \Delta \theta_e$= $\pm$ 1.6~$^\circ$ 
and $ \Delta \phi_e$= $\pm$ 3.4~$^\circ$.
For $Q^2$= 0.1 (0.2) ${\rm GeV}^2/c^2$ or values close to it the virtual photon 
energies were chosen between 30--90 MeV (80--140 MeV) with a central 
value of 60 MeV 
(110 MeV). Each of these two $\omega$-ranges were divided into 7 bins 
of length 10 MeV.

The three-fold cross section for inclusive scattering has the well known 
form (Eq.~(I.3))
\begin{equation}
{ {d^{\, 3} \sigma} \over { d \hat{k}' \, d k_0' } } \ = \
\sigma_{\rm Mott} \, 
\left\{ v_L R^L + v_T R^T 
\, + \, h \left( v_{TL'} R^{TL'} + v_{T'} R^{T'} \right) \right\} .
\label{eq:sigma3}
\end{equation}
Here $\sigma_{\rm Mott}$ is the Mott cross section, $v_i$ are analytically 
known kinematical factors and the $R^i$ 
are inclusive response functions divided into two groups. The first one 
is present 
for unpolarized electrons, the second one goes with the helicity $h$ 
of the electron beam. 
The primed response functions also depend on the orientation of the $^3$He 
spin in relation to the photon direction (see Eqs.~(I.56-57)). The corresponding angles 
are denoted by $ \theta^\star $ and $\phi^\star$.
The cross section in Eq.~(\ref{eq:sigma3}) was averaged over the 10 MeV wide
$\omega$-bins and over the angular spread around the central electron scattering
angles. In order to perform the averaging a sufficiently fine grid in
$\theta_e$ and $\omega$ has been chosen for which the four response functions 
have been calculated. This required quite a few hundred solutions of 
the corresponding Faddeev equations. 
The actual averaging was performed via a Monte Carlo procedure based on the response
functions known on the grid of electron angles and electron energies. This Monte Carlo
procedure takes into account the finite momentum and angular acceptance 
of the experiment.

The asymmetry is defined as
\begin{equation}
A \ = \ 
{
{
\left. { {d^{\, 3} \sigma} \over { d \hat{k}' \, d k_0' } }\right|_{h=1}
\ - \ \left. { {d^{\, 3} \sigma} \over { d \hat{k}' \, d k_0' } }\right|_{h=-1}
} \over
{
\left. { {d^{\, 3} \sigma} \over { d \hat{k}' \, d k_0' } }\right|_{h=1}
\ + \ \left. { {d^{\, 3} \sigma} \over { d \hat{k}' \, d k_0' } }\right|_{h=-1}
} 
} .
\label{eq:AN}
\end{equation}
This ratio is  formed out of the averaged cross sections. 
As a consequence one arrives at (see Eq.~(I.58))
\begin{equation}
A_{average} \ = \
{
{
\int \, d \Omega \, \sigma_{\rm Mott} \, \left\{ v_{T'} \tilde{R}^{T'} \cos \theta^\star 
\, + \, v_{TL'} \tilde{R}^{TL'} \sin \theta^\star \cos \phi^\star \right\} 
} \over
{
\int \, d \Omega \, \sigma_{\rm Mott} \, \left\{ v_{L} R^{L}
\, + \, v_{T} R^{T} \right\} 
} 
} \ \equiv \ { \Delta \over \Sigma } ,
\label{eq:ANav}
\end{equation}
where $d \Omega $ stands for the averaging. (We factored off the 
$\theta^\star \, \phi^\star $-dependence introducing the
response functions with tilde). For $\theta^\star$= 0~$^\circ$ or close to it 
one focuses on  $\tilde{R}^{T'} $ and a corresponding $ A_{T'}$, which 
in a plane wave impulse approximation (PWIA) is essentially proportional 
to $\left( G_M^n \right)^2$ (see Eq.~(I.77)). 
In the actual experiment one has to live with 
$\theta^\star \le $ 10~$^\circ$ (7.8~$^\circ$) 
for $Q^2$= 0.1 (0.2) ${\rm GeV}^2/c^2$ and the  corresponding $\phi^\star$ 
 is close to 0~$^\circ$ or 180~$^\circ$.

The searched for magnetic form factor of the neutron was parameterized as
\begin{equation}
G_M^n \left( Q^2 \right) \ \equiv \ \lambda  \, 
\left. G_M^n \left( Q^2 \right) \right|_{\rm model} ,
\label{eq:GMNparam}
\end{equation}
where  $\left. G_M^n \left( Q^2 \right) \right|_{\rm model} $ was taken from~\cite{Hoehler}. 
In order  to keep the computer time below an acceptable limit the
averaging process was performed only for $\lambda$= 1. 
For the $\lambda$-values in the neighborhood of $1$ it was assumed that the change  
for $A_{T'}$ from point geometry (fixed $\omega$ and central electron angles) 
to the averaged case is the same as for $\lambda$= 1. Because of
the smallness of the $\lambda$ interval around $\lambda$= 1 (see below) this 
is highly plausible. In this manner one generated for each $\lambda$-value 
theoretical $A_{T'}$-values according to the seven $\omega$-bins. 
The final step is the adjustment of the magnetic form factor of the neutron, 
$G_M^n $. Out of the seven $\omega$-bins three central values
in the QFS region were selected (for $Q^2$= 0.1 ${\rm GeV}^2/c^2$
$\omega$= 50, 60, 70 MeV and for $Q^2$= 0.2 ${\rm GeV}^2/c^2$
$\omega$= 100, 110, 120 MeV) and an additional averaging was performed
\begin{equation}
\bar{A} \ \equiv \
{ { \sum_{j=1}^3 A_j \Sigma_j } \over { \sum_{i=1}^3 \Sigma_i } } \ = \  
\sum_{j=1}^3 { \Delta_j \over \Sigma_j } \, { \Sigma_j \over {\sum_{i=1}^3 \Sigma_i} }
\ \equiv \ \sum_{j=1}^3 { \Delta_j \over \Sigma_j } \, w_j
\ = \ \sum_{j=1}^3 A_j \, w_j
\label{eq:Addav}
\end{equation}
The indices $i$ and $j$ refer to the three experimental bins. 
In the third equality weight factors $w_j$ are introduced, 
which in the actual performance were taken from the experiment 
(counts related to the unpolarized cross section).

\begin{figure}[h!]
\epsfbox{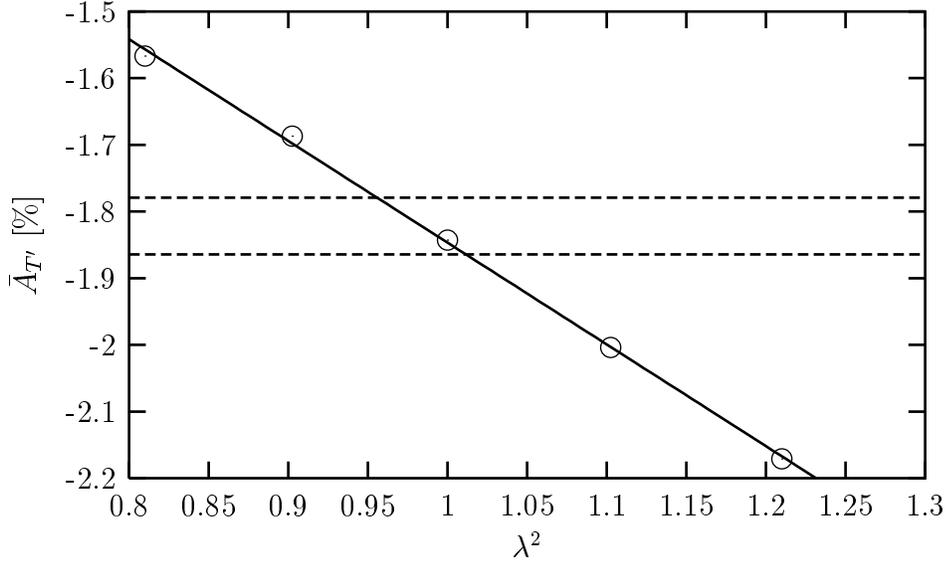}
\caption[ ]{The averaged asymmetry $\bar{A}_{T'}$ of Eq.~(\protect\ref{eq:Addav})
            around the quasielastic peak against the $ \lambda^2$-factor
            for $Q^2$= 0.1 ${\rm GeV}^2/c^2$. The solid curve is a result of a fit.
            Dashed curves show the experimental bounds for $A_{T'}$.}
\label{figq1lambda}
\end{figure}

\begin{figure}[h!]
\epsfbox{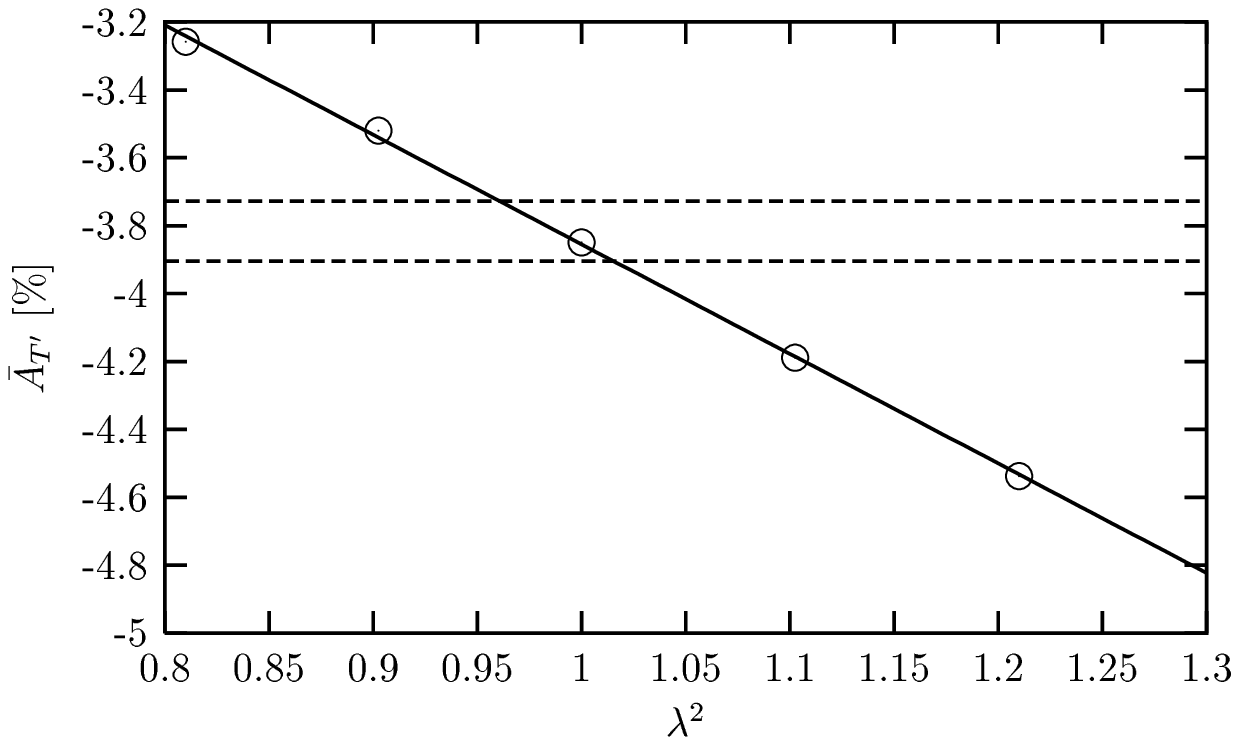}
\caption[ ]{The same as in Fig.~\protect\ref{figq1lambda} for $Q^2$= 0.2 ${\rm GeV}^2/c^2$.}
\label{figq2lambda}
\end{figure}

In this manner one arrives at the  $\lambda^2$-dependence of $\bar{A}$, 
which turned out to be rather close to a straight line. 
This is depicted in Figs.~\protect\ref{figq1lambda} and \protect\ref{figq2lambda}
for $Q^2$= 0.1 and 0.2 ${\rm GeV}^2/c^2$, respectively, together with the
experimental values. One reads off the $\lambda$-values leading 
to $G_M^n $ values as given in Table~\protect\ref{gmn_tbl}.
There $G_M^n (dipole) =\mu_n / (1 + Q^2/0.71)^2 $, where $\mu_n$ is the 
magnetic moment of the neutron.

\begin{figure}[h!]
\epsfbox{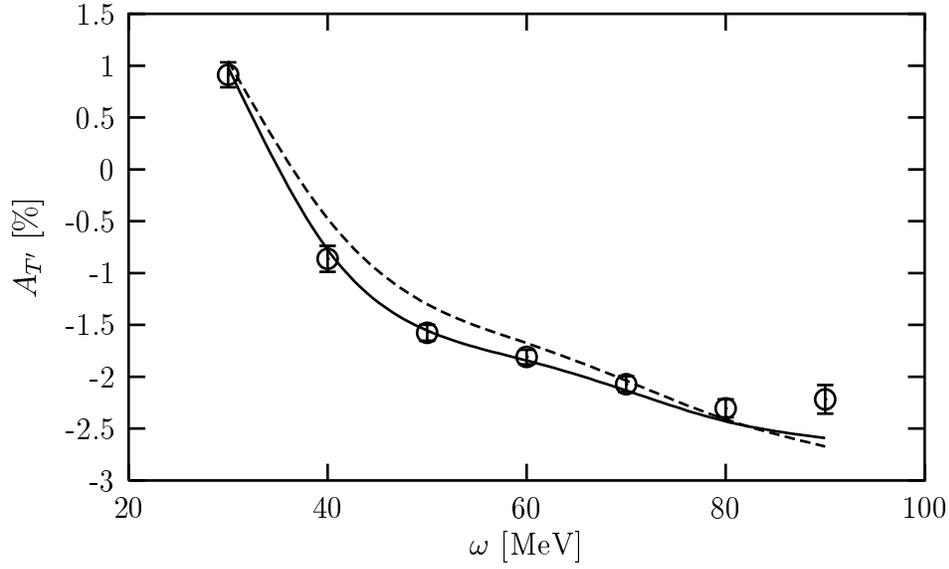}
\caption[ ]{The asymmetry $A_{T'}$ against the energy transfer $\omega$
            for $Q^2$= 0.1 ${\rm GeV}^2/c^2$. The curves describe full (averaged) 
            Bonn B predictions with the single nucleon current (dashed)
            and with the single nucleon current plus the $\pi$- and $\rho$-MEC (solid).
            Data are from~\protect\cite{doublestar}.} 
\label{figq1exp}
\end{figure}

\begin{figure}[h!]
\epsfbox{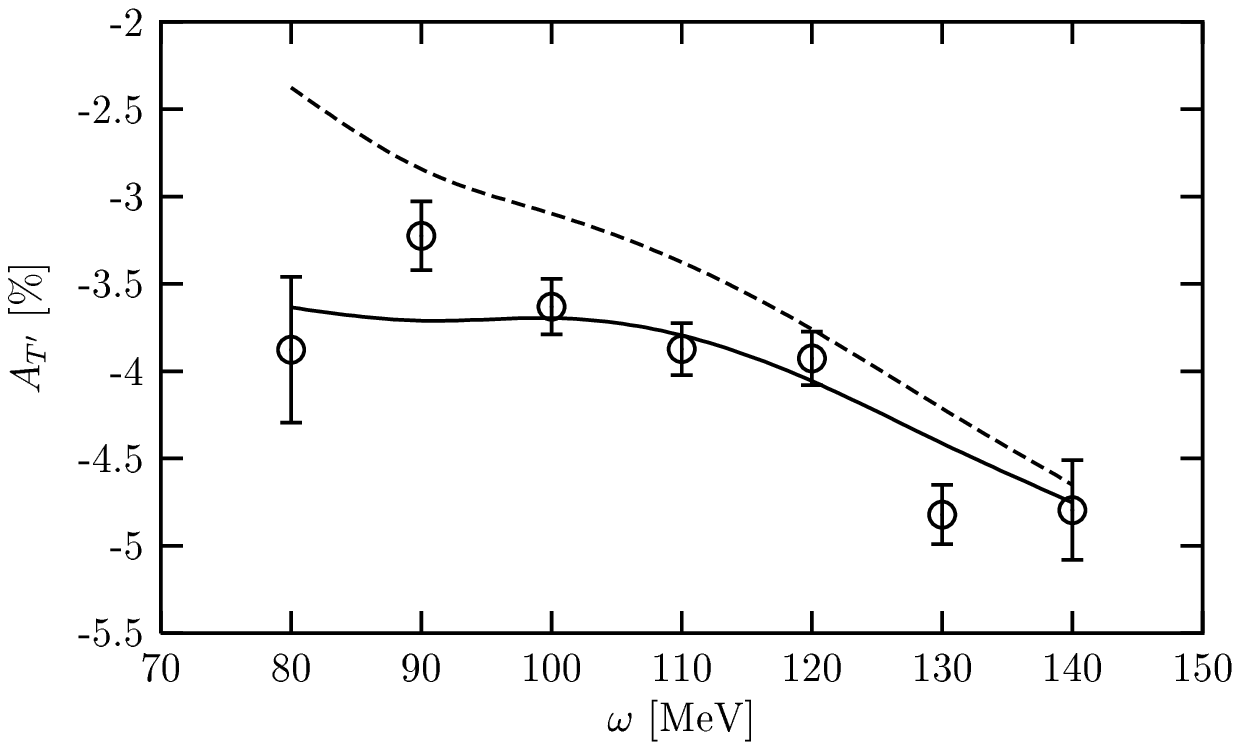}
\caption[ ]{The same as in Fig.~\protect\ref{figq1exp} for $Q^2$= 0.2 ${\rm GeV}^2/c^2$.}
\label{figq2exp}
\end{figure}

\begin{figure}[h!]
\epsfbox{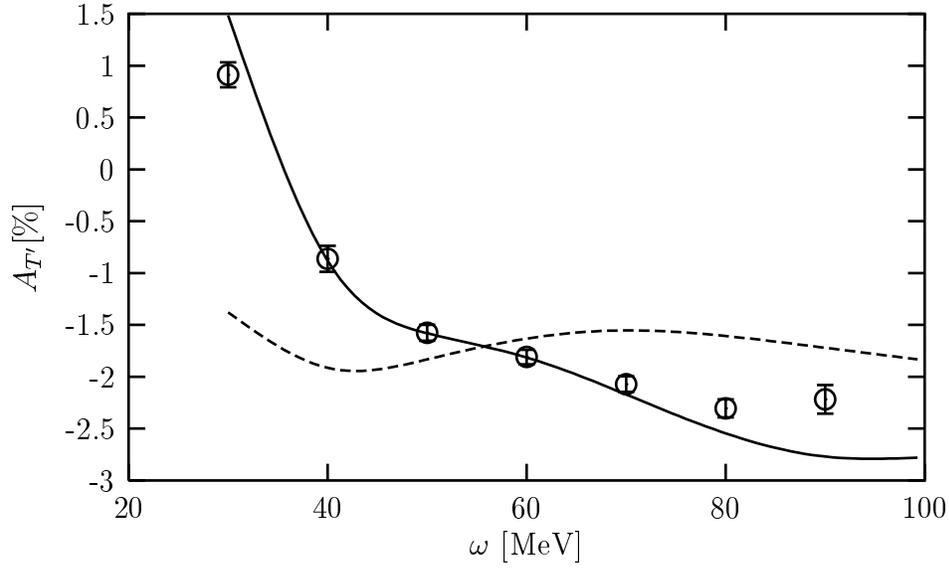}
\caption[ ]{The asymmetry $A_{T'}$ against the energy transfer $\omega$
            for $Q^2$= 0.1 ${\rm GeV}^2/c^2$. The curves describe point geometry 
            results obtained with the AV18 potential: PWIA (dashed) 
            and the full prediction (solid). In both cases 
            the single nucleon current plus the $\pi$- and $\rho$-MEC is used.
            Data are from~\protect\cite{doublestar}.} 
\label{figq1pwia}
\end{figure}

\begin{figure}[h!]
\epsfbox{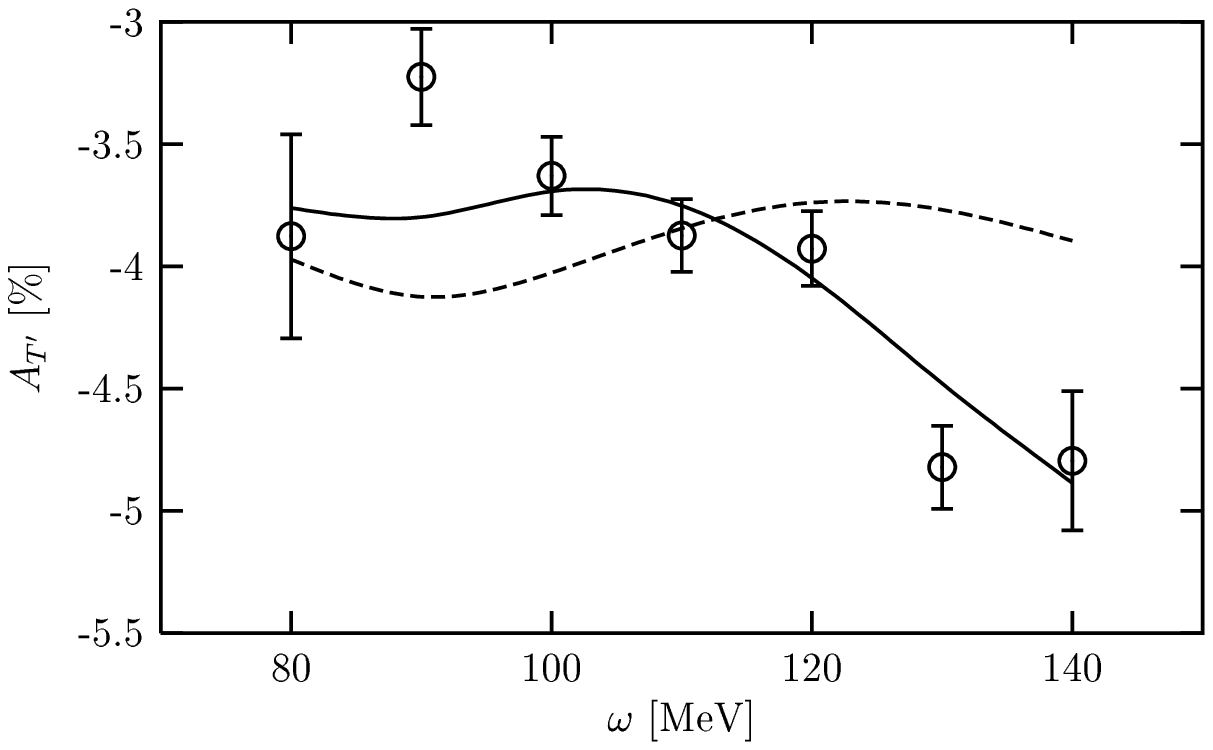}
\caption[ ]{The same as in Fig.~\protect\ref{figq1pwia} for $Q^2$= 0.2 ${\rm GeV}^2/c^2$.}
\label{figq2pwia}
\end{figure}

Having adjusted $G_M^n $ we can display the $\omega$-dependence 
of $A_{T'}$ in Figs.~\protect\ref{figq1exp} and \protect\ref{figq2exp}
in comparison to the experimental values. We see an essentially perfect 
agreement between theory and experiment.
We also show the theoretical result without MEC's but including 
the full final state interaction. 
Clearly the MEC's provide an important shift and should be not neglected. 
Also the final state interaction itself plays a very important role 
since the PWIA result (see Figs.~\protect\ref{figq1pwia} and~\protect\ref{figq2pwia})
is far off. Note our PWIA neglects all final state interactions.

\begin{figure}[h!]
\epsfbox{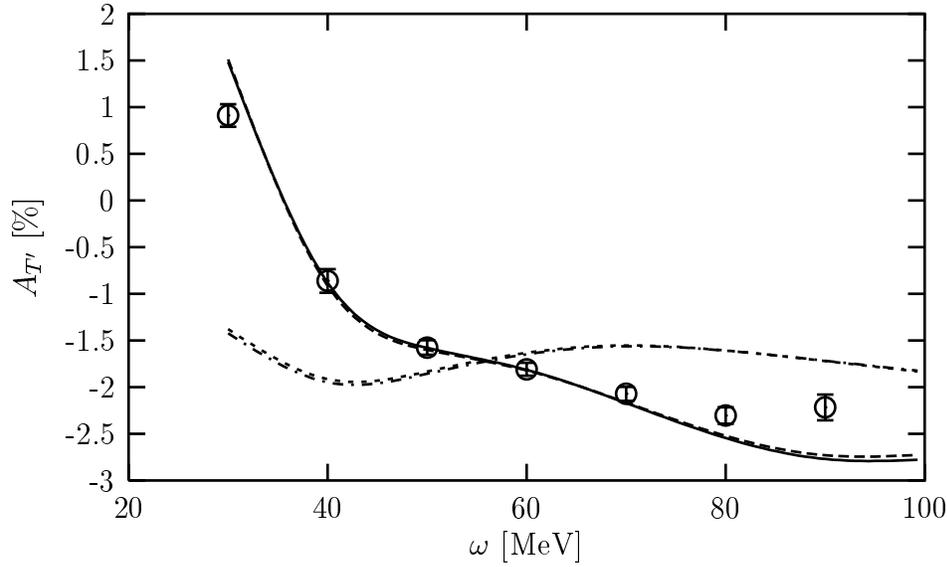}
\caption[ ]{The asymmetry $A_{T'}$ against the energy transfer $\omega$
            for $Q^2$= 0.1 ${\rm GeV}^2/c^2$. The curves describe PWIA
            point geometry results with $G_E^n = 0$ (dashed-dotted),
            and $G_E^n \ne 0$ (dotted),
            full point geometry results with $G_E^n = 0$ (dashed)
            and $G_E^n \ne 0$ (solid).
            All results are obtained with the AV18 potential. 
            The single nucleon current plus the $\pi$- and $\rho$-MEC is used.
            Data are from~\protect\cite{doublestar}.} 
\label{figq1gen}
\end{figure}

\begin{figure}[h!]
\epsfbox{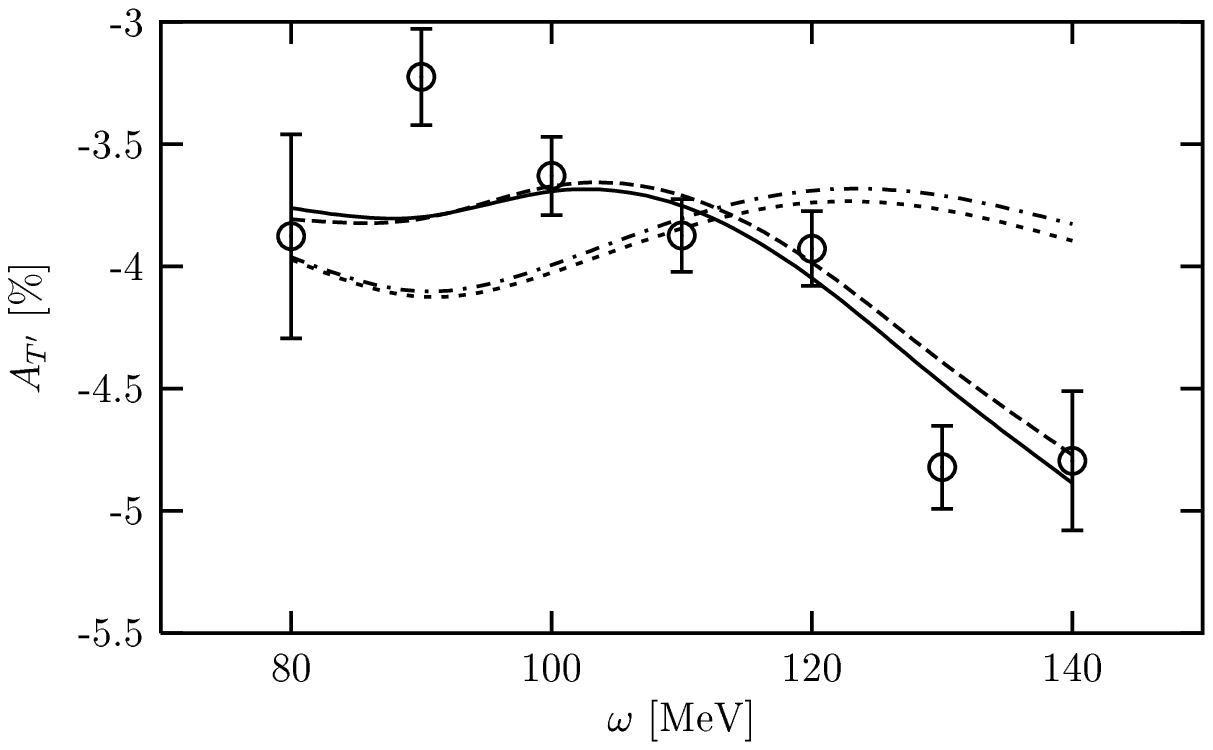}
\caption[ ]{The same as in Fig.~\protect\ref{figq1gen} for $Q^2$= 0.2 ${\rm GeV}^2/c^2$.}
\label{figq2gen}
\end{figure}

The electric form factor of the neutron, $G_E^n$, is not yet very well known  
(see  section~\ref{secIII}), but enters into our calculation. Its effect is 
totally negligible as can be seen in Figs.~\protect\ref{figq1gen} and \protect\ref{figq2gen}. 
There we compare $A_{T'}$ (for point geometry) evaluated with 
$G_E^n$ according to~\cite{Hoehler} and putting it to zero.

\begin{figure}[h!]
\epsfbox{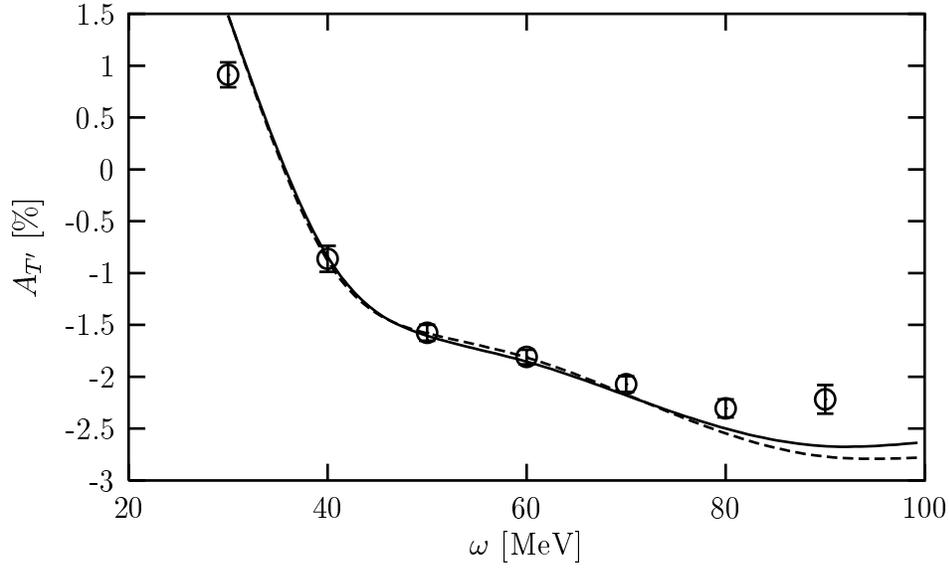}
\caption[ ]{The asymmetry $A_{T'}$ against the energy transfer $\omega$
            for $Q^2$= 0.1 ${\rm GeV}^2/c^2$. The curves describe full point geometry 
            results obtained with the AV18 potential (dashed) 
            and with the Bonn B potential (solid). In both cases 
            the single nucleon current plus the $\pi$- and $\rho$-MEC are used.
            Data are from~\protect\cite{doublestar}.} 
\label{figq1pot}
\end{figure}

\begin{figure}[h!]
\epsfbox{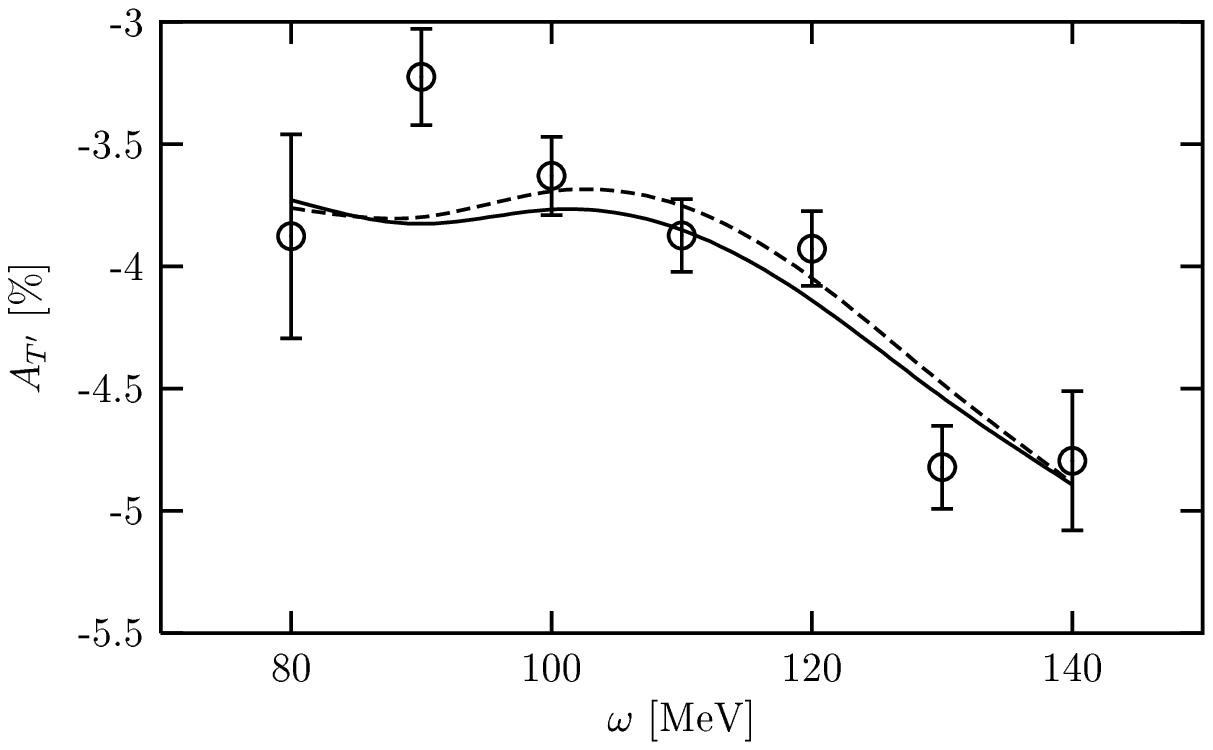}
\caption[ ]{The same as in Fig.~\protect\ref{figq1pot} for $Q^2$= 0.2 ${\rm GeV}^2/c^2$.}
\label{figq2pot}
\end{figure}

For point geometry we also performed full fledged 
calculations based on the AV18 NN force~\cite{AV18} 
and the $\pi$- and $\rho$-like exchange currents
according to the Riska prescription~\cite{Riska}.
Both calculations, for Bonn~B and AV18 agree very well as shown 
in Figs.~\protect\ref{figq1pot} 
and \protect\ref{figq2pot}.

Finally we mention that we also included $\pi$- and $\rho$-exchange 
currents with an intermediate $\Delta$  (in a static approximation).  
Their effects was very small and lead to an estimated change of $G_M^n$
by less than 2~{\%}.

The data in Figs.~\protect\ref{figq1exp}-\protect\ref{figq2pot} 
are radiatively corrected. This thorough 
procedure was performed with the help of numerous full fledged Faddeev 
calculations and will be described in a separate article~\cite{Feng}.

\begin{figure}[h!]
\epsfbox{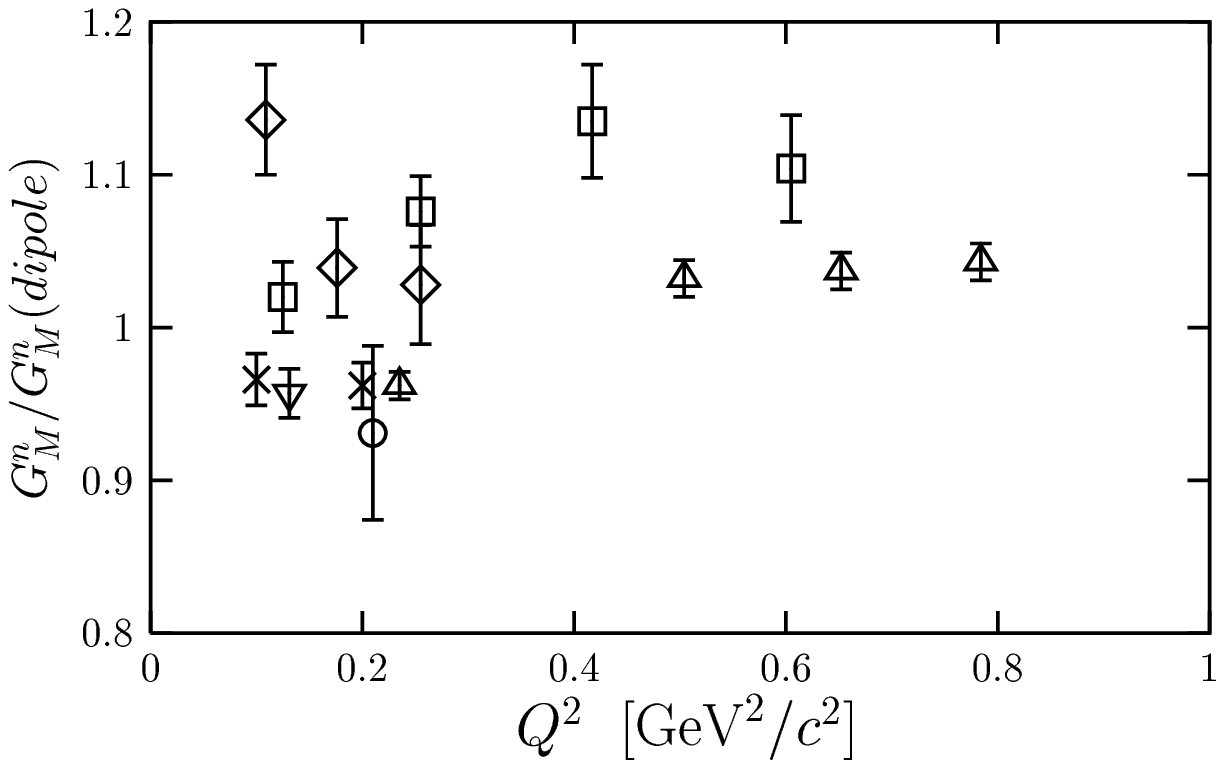}
\caption[ ]{$G_M^n$-values extracted from different measurements
on the deuteron 
(\protect\cite{Markowitz93} ($\Diamond$),
 \protect\cite{Anklin94} ($\bigtriangledown$),
 \protect\cite{Bruins95} ($\Box$),
 \protect\cite{Anklin98} ($\bigtriangleup$))
and on $^3$He
(\protect\cite{Gao94} ($\bigcirc$),
 \protect\cite{doublestar} ($\times$)).}
\label{Gmnothers}
\end{figure}

Our $G_M^n$ value extracted from that inclusive experiment on $^3$He 
agrees very well with results achieved in recent experiments on the 
deuteron~\cite{Anklin94,Anklin98}. This is shown in Fig.~\protect\ref{Gmnothers}
together with other data.

We refrained from a theoretical analysis of data taken in the same experiment 
at higher $Q^2$-values~\cite{doublestar}, since one has to expect that 
relativity will play a non-negligible role.
This is left to a future investigation.

\section{Extraction of the electric form factor of the neutron}
\label{secIII}

\begin{figure}[h!]
\centerline{\mbox{\epsfysize=100mm \epsffile{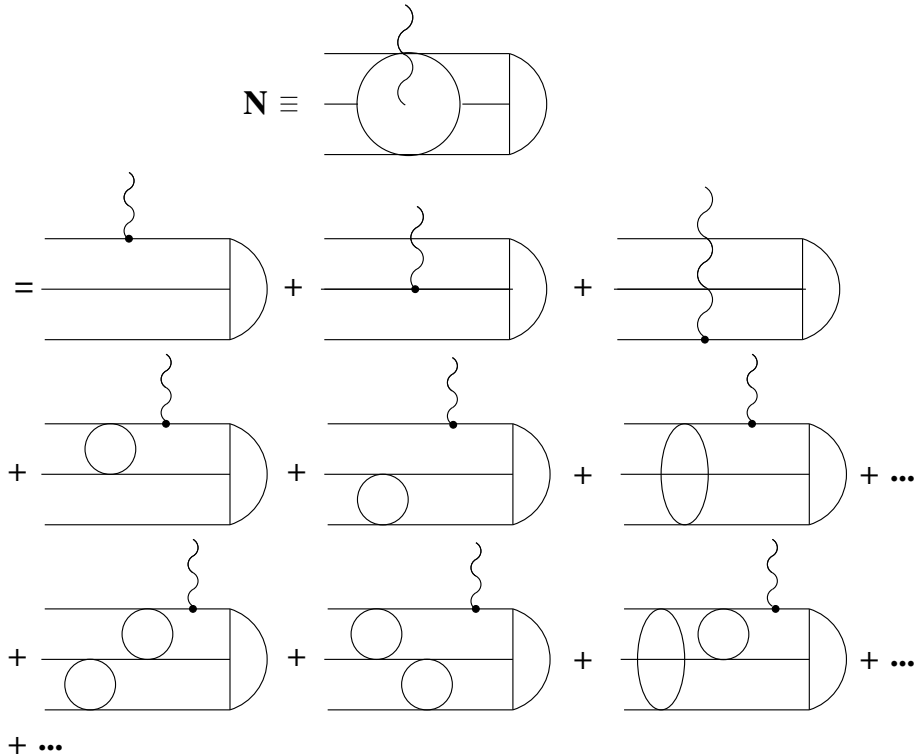}}}
\vspace{0.5cm}
\caption{The multiple rescattering series for the process $^3$He(e,e'n). 
The half moon stands for the $^3$He state, the wavy line for the photon,
horizontal lines for freely propagating nucleons and the ovals 
for NN t-matrices. The dots in the third line stand for processes, where
the photon is absorbed on the other two nucleons.}
\label{diagram}
\end{figure}

In this section we shall analyze a measurement of 
$\overrightarrow{^3{\rm He}}(\vec{e},e'n)$ 
carried through at MAMI~\cite{triplestar} with the aim to extract $G_E^n$. 
Our theoretical formalism has been
described in~\cite{ref7}. Nevertheless to clearly shed light onto the reactions
going on after the virtual
photon has been absorbed we would like to lay out the multiple rescatterings
and their summation into a Faddeev like integral equation. In the
literature erroneously often just the very first few terms
are taken into account.
In a graphical representation the full photon-induced break-up process
is an infinite
sum of the type shown in Fig.~\protect\ref{diagram}.
We assumed the absorption 
of the photon on a single nucleon.
Obviously the diagrams can be generalized by photon absorption processes 
on two or three nucleons.
This infinite sequence of processes has its algebraic
counterparts

\begin{eqnarray}
N &=& \left( j(1) + j(2) + j(3) \right) \, \vert \Psi_{^3He}\rangle \cr 
  &+& \left( t_{12} +  t_{23} +  t_{31} \right) \, G_0 \, 
      \left( j(1) + j(2) + j(3) \right) \, \vert \Psi_{^3He}\rangle \cr
  &+& \sum_{i=1}^3 \, \sum_{k < l} \, \sum_{m < n \ne k < l} \
       t_{mn} \, G_0 \, t_{kl} \, G_0 \, j(i) \, \vert \Psi_{^3He}\rangle \, + \, \cdots 
\label{eq:N}
\end{eqnarray}

It is convenient to introduce the notation $t_{ij} \equiv t_k$ 
($ijk$ = $123$ etc.) and $P \equiv P_{12} P_{23} + P_{13} P_{23} $.
Then it requires little
work to put Eq.~(\ref{eq:N}) into the form
\begin{eqnarray}
N &=& \left( 1 + P \right) \, j(1) \, \vert \Psi_{^3He}\rangle \cr 
  &+& \left( 1 + P \right) \, t_{1} \, G_0 \, 
      \left( 1 + P \right) \, j(1) \, \vert \Psi_{^3He}\rangle \cr     
  &+& \left( 1 + P \right) \, t_{1} \, G_0 \, P \, 
      t_{1} \, G_0 \, \left( 1 + P \right) \, j(1) \, \vert \Psi_{^3He}\rangle 
      \, + \, \cdots
\label{eq:N2}
\end{eqnarray}
The first term contains no final state interaction and we split off,
what we shall call plane wave impulse approximation (PWIA),
\begin{equation}
N^{\rm PWIA} \equiv j(1) \, \vert \Psi_{^3He}\rangle 
\end{equation}
and call the whole term 
\begin{equation}
N^{\rm PWIAS} \equiv \left( 1 + P \right) \, j(1) \, \vert \Psi_{^3He}\rangle ,
\end{equation}
where S stands for full antisymmetrisation.
All the additional terms contain rescattering contributions of
increasing oder in the NN $t$-operator:
\begin{eqnarray}
N^{\rm rescatt} &\equiv& \left( 1 + P \right) \, 
 [ t_{1} \, G_0 \, + t_{1} \, G_0 \, P \, t_{1} \, G_0 + \cdots ]
   \, \left( 1 + P \right) \, j(1) \, \vert \Psi_{^3He}\rangle \cr
  &\equiv& \left( 1 + P \right) \, \vert U \rangle .
\label{eq:U}
\end{eqnarray}
A more general inspection reveals that
\begin{eqnarray}
\vert U \rangle \ = \ 
\left( 1 \ + \  t_{1} \, G_0 \, P \  + \ t_{1} \, G_0 \, P \, t_{1} \, G_0 \, P 
+ \cdots \right) \, 
  t_{1} \, G_0 \, \left( 1 + P \right) \, j(1) \, \vert \Psi_{^3He}\rangle ,
\label{eq:U2}
\end{eqnarray}
where inside the bracket the operator sequence 
$ t_{1} \, G_0 \, P $ occurs in
increasing powers. As an immediate consequence one derives
\begin{eqnarray}
\vert U \rangle \ = 
\ t_{1} \, G_0 \, \left( 1 + P \right) \, j(1) \, \vert \Psi_{^3He}\rangle
\ + \ t_{1} \, G_0 \, P \, \vert U \rangle ,
\label{eq:U3}
\end{eqnarray}
which is the central integral equation for the amplitude $ \vert U \rangle$. 
Via Eq.~(\ref{eq:U}) 
it provides the whole rescattering
amplitude (up to the symmetrisation 
$ \left( 1 + P \right) $).
This integral equation is
of the Faddeev type because of the typical Faddeev structure of its
kernel.
The same kernel occurs for 3N scattering processes~\cite{ref9}, 
only the driving term is different there.

An often used approximation for quasielastic processes in the literature is
\begin{eqnarray}
N \approx j(1) \, \vert \Psi_{^3He}\rangle \ + \
t_{23} \, G_0 \, j(1) \, \vert \Psi_{^3He}\rangle \ \equiv \ N^{\rm FSI23} .
\label{eq:FSI23}
\end{eqnarray}
Here antisymmetrisation in the final state is neglected and one 
rescattering in the NN
$t$-operator $t_{23}$ is only allowed for the two spectator nucleons 
(which do not absorb the photon). This amplitude is also sometimes 
called PWIA, but not in this article.

Needless to say that the full amplitude, now supplemented by the
proper vector indices for the current operator is identical
to the standard form of the nuclear matrix element
\begin{eqnarray}
N^\mu \ = \ \langle \Psi_f^{(-)} \vert 
\sum_i \, j^\mu (i) \vert \Psi_{^3He}\rangle .
\label{eq:Nmu}
\end{eqnarray}
We refer to~\cite{ref7} for the verification.

The sixfold differential cross section for the exclusive process under
discussion has the well known form~\cite{Donnelly}
\[
{ {d^{\, 6} \sigma} \over { d \hat{k}' \, d k_0' \, d \hat{p}_n d p_n} } \ = \
\sigma_{\rm Mott} \, p_n^2 \, { { p \, m_N} \over 2 }
\]
\begin{eqnarray}
\times \int d \hat{p} 
\left\{ v_L R^L + v_T R^T + v_{TT} R^{TT} + v_{TL} R^{TL}
\, + \, h \left( v_{TL'} R^{TL'} + v_{T'} R^{T'} \right) \right\} .
\label{eq:sigma6}
\end{eqnarray}
Here 
$ \hat{k}' $, $k_0'$, $\hat{p}_n$, $p_n$, $p$, $\hat{p}$
in turn are the unit vector in the direction of the scattered
electron, its energy, the unit vector of the knocked out neutron, its
momentum, the magnitude of the relative momentum of the undetected
two protons and finally the unit vector pointing into the direction
of that relative momentum.

Throughout this article we use a strictly nonrelativistic notation.

The information on $G_E^n$ magnified by the product with $G_M^n$ is contained
in $R_{TL'}$ as can be explicitely
seen working out PWIA~\cite{Ziemer} (see also~\cite{triplestar}). In order to isolate the primed
structure functions one forms an asymmetry $A$ of the cross section
 with respect to the electron
helicities $h = \pm 1$.
One finds the well known result
\begin{eqnarray}
A \ = \
{ { \int d \hat{p} \left( v_{TL'} R^{TL'} + v_{T'} R^{T'} \right) } \over 
{ \int d \hat{p} \left( v_L R^L + v_T R^T + v_{TT} R^{TT} + v_{TL} R^{TL} \right) } }
\label{eq:Asymmetry}
\end{eqnarray}

In the experiment two perpendicular polarization axis for the spin  of
$^3$He have been chosen, $\vec{S}_\parallel$  
and $\vec{S}_\perp$, leading to $A_\parallel$  and $A_\perp$. 
In an optimal set up  $\vec{S}_\parallel$ and $\vec{S}_\perp$ 
would be parallel and perpendicular to the direction
of the virtual photon. Then it is well known from the expression for PWIA
that under this approximation
\begin{eqnarray}
{ A_\perp \over A_\parallel } \propto { G_E^n \over G_M^n }
\label{eq:Aratio}
\end{eqnarray}
Therefore the aim will be to extract $G_E^n$ from the measured value of
that ratio. This is under the assumption that $G_M^n$ is sufficiently
well known. Our contribution in this article is to show that the final state interaction (FSI) 
does not wash out the signal for $G_E^n$ and that taking FSI into account  
is crucial.

\begin{figure}[h!]
\epsfbox{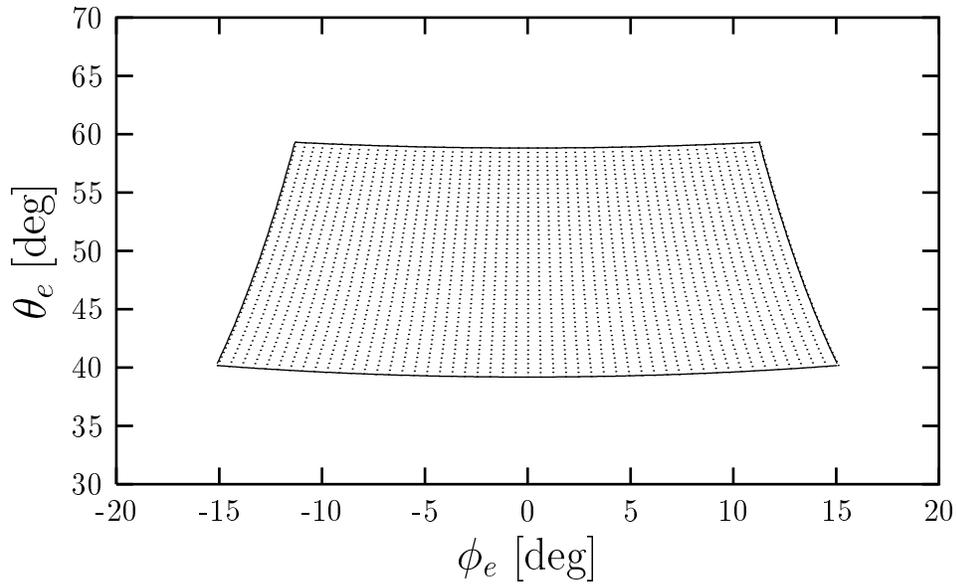}
\caption[ ]{Angular acceptance of the electron detector.}
\label{fig721}
\end{figure}

\begin{figure}[h!]
\epsfbox{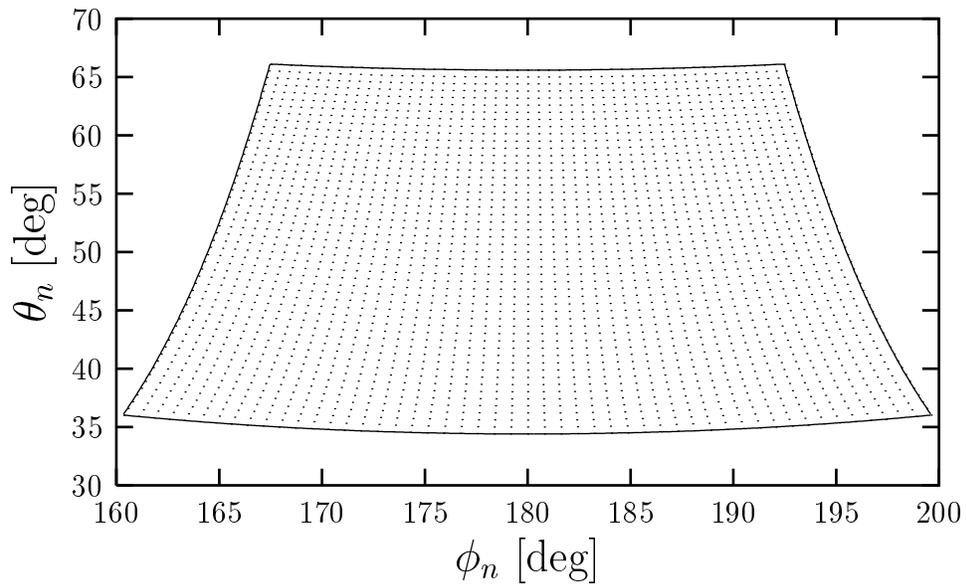}
\caption[ ]{Angular acceptance of the neutron detector.}
\label{fig722}
\end{figure}

\begin{figure}[h!]
\epsfbox{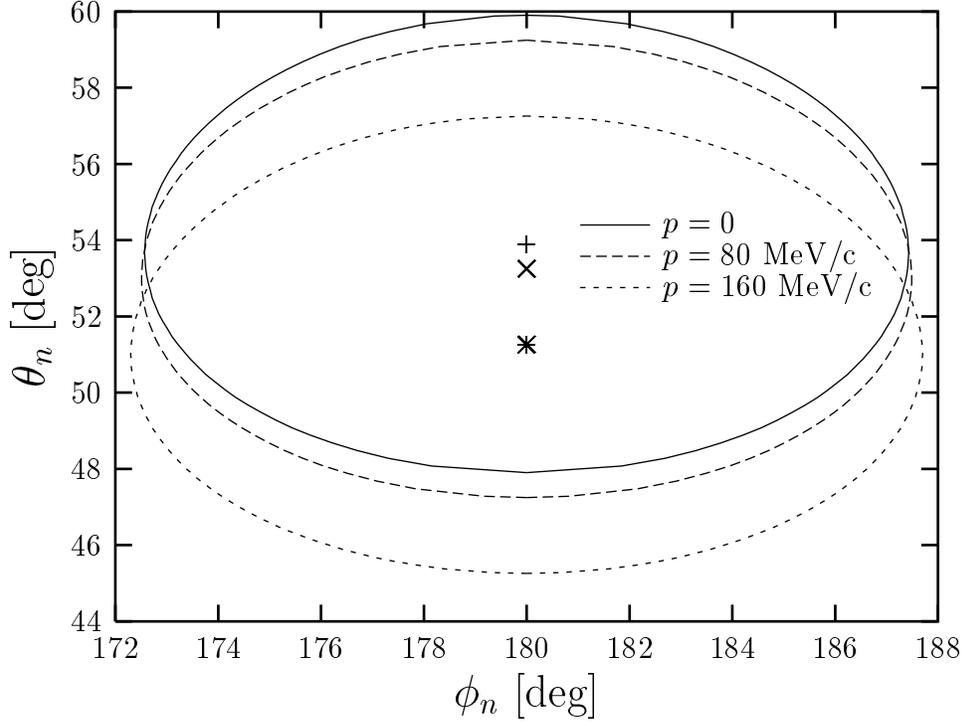}
\caption[ ]{Integration limits for $\hat{p}_n$ (solid curve) for the example 
$\theta_e$ = 43 $^\circ$, $\phi_e$ = 0 $^\circ$ and $p_n$ = 530 MeV/c together with the 
direction of the photon $+$. The dashed and dotted curves are for $p$ = 80 and 160 MeV/c
and the corresponding directions of the photon are given by $\times$ and $*$.}
\label{fig73}
\end{figure}

We shall now describe the experimental conditions under which the data 
were taken. The electron- and neutron-detectors covered 
a wide range of angles as displayed 
in Figs.~\protect\ref{fig721}-\protect\ref{fig722}.
Depending
on the electron scattering angle $\theta_e$ only neutron momenta
within certain cuts were accepted as shown 
in Table~\protect\ref{tab71}.
Since the energy of the scattered electron was not measured (except for
excluding pion-production)
the direction  of the virtual photon was not known. However it was
possible to correlate the directions
of the photon and the knocked out neutron in the following manner~\cite{BeckerPri}.
Take the relative momentum $p$ of the two undetected
protons to be zero. Then for fixed values of $\hat{k}'$, $\hat{p}_n$
and $p_n$, $k_0'$ and therefore $\vec Q$,
the virtual photon momentum, follow kinematically. Now only those angles
   were allowed such that the angle between $\vec Q$ and $\hat{p}_n$ 
was smaller or equal to 6~$^\circ$.
In our theoretical analysis we also took
that constraint into account. As an example we show 
in Fig.~\protect\ref{fig73}
the allowed region for the neutron angles for given values of 
$\theta_e$ = 43~$^\circ$, $\phi_e$ = 0~$^\circ$, $p_n$ = 530 MeV/c and $p$ = 0 (solid curve). 
In reality there
is a distribution of $p$-values, see Fig.~\protect\ref{fig16} below. 
Consequently for that region 
of neutron angles also other events with $p \ne 0$ contribute, which belong
to different directions of $\vec Q$. This is also shown
in Fig.~\protect\ref{fig73}. In the
 worst case the angle
between $\hat{Q}$  and $\hat{p}_n$ can be as large as 9~$^\circ$
for the tails of the $p$-distribution.

Since the energy of the scattered electron has not been measured the
 cross section reduces to a five-fold one
\[
{ {d^{\, 5} \sigma} \over { d \hat{k}' \, d \hat{p}_n d p_n} } \ = \
{ m_N \over 2} \, \sigma_{\rm Mott} \, p_n^2 \, 
\]
\begin{eqnarray}
\times 
\int d {k'}_0 \, p \,
\int d \hat{p} 
\left\{ v_L R^L + v_T R^T + v_{TT} R^{TT} + v_{TL} R^{TL}
\, + \, h \left( v_{TL'} R^{TL'} + v_{T'} R^{T'} \right) \right\} .
\label{eq:sigma5}
\end{eqnarray}
We convert the ${k'}_0$ integration into one over $p$, take into account the
 experimental acceptances and end up
with the summed up cross section
\[
\Delta \sigma \, \equiv \, \int_{ \Delta \hat{k}' } \, d \hat{k}' \, 
\sigma_{\rm Mott} \,
\int_{ \Delta p_n }  d p_n p_n^2 \,
\int_{ \Delta \hat{p}_n } d\hat{p}_n \, \int d \vec p \, \rho 
\]
\[
\times 
\left\{ v_L R^L + v_T R^T + v_{TT} R^{TT} + v_{TL} R^{TL}
\, + \, h \left( v_{TL'} R^{TL'} + v_{T'} R^{T'} \right) \right\} .
\]
\begin{eqnarray}
\equiv \ \int d \Omega 
\left\{ v_L R^L + v_T R^T + v_{TT} R^{TT} + v_{TL} R^{TL}
\, + \, h \left( v_{TL'} R^{TL'} + v_{T'} R^{T'} \right) \right\} .
\label{eq:sigmaOmega}
\end{eqnarray}

In our nonrelativistic formulation  $\rho$  has the form
\begin{eqnarray}
\rho \ = \
\frac{2\, m_N \, p^2}{ \sqrt{
\left( k \cos \theta_e - \vec{p}_n \cdot \hat{k}' -2 m_N \right)^2 
+ 4 m_N \left( k + \epsilon_{^3He} \right) - 4 p^2 - 3 p_n^2 - k^2 
+ 2 \vec{p}_n \cdot \vec{k} } }
\label{eq:rho}
\end{eqnarray}

Since we shall form asymmetries there is no need to determine the
value of the covered phase space. Using 
$\Delta \sigma $ for the two $^3$He-spin directions one can form
the two asymmetries $A_{\perp}, A_\parallel$ for the  
$\perp$  and $\parallel$ orientations of the $^3$He spin and finally their ratio $V$
\begin{eqnarray}
V \ \equiv \ { A_\perp \over A_\parallel } \ = \
{ { \int d \Omega \left( v_{TL'} R^{TL'} + v_{T'} R^{T'} \right)_\perp } \over 
{ \int d \Omega \left( v_{TL'} R^{TL'} + v_{T'} R^{T'} \right)_\parallel } } \, \cdot \, 1 .
\label{eq:V}
\end{eqnarray}
The ``1'' in Eq.~(\ref{eq:V}) 
denotes the corresponding ratio for the helicity independent
parts of $\Delta\sigma$. It turned out that this latter ratio was extremely close 
to 1 (within less than 0.1 \%).

Before we shall present our results for $V$ as a function of  a parametrisation
of $G_E^n$ we would like to give
some insight into the functions entering Eq.~(\ref{eq:V}). 
For some fixed directions
of $ \hat{k}' $ and $ \hat{p}_n$  and some value of $p_n$  
contained in the domain $\Omega$  we define
the quantities 
\begin{eqnarray}
\Lambda_\beta^\alpha \, \equiv \, \int d \hat{p} \, R_\beta^\alpha,
\label{eq:Lambda}
\end{eqnarray}
which depend on $p$. For $\alpha$ we choose  $T'$ and $TL'$
and $\beta$ corresponds to $\perp$   and $\parallel$-orientations of the $^3$He spin. 
Fig.~\protect\ref{fig16}  tells
 us that indeed the distribution of the $p$-values
is peaked at low values, where the maxima occur at kinetic energy values
 of relative motion of the two protons of about 0.4 MeV. At around 140 MeV/c
 the $p$-distribution has essentially vanished.
Those curves in Fig.~\protect\ref{fig16} refer to full FSI. 
In contrast the corresponding curves for PWIA, displayed in Fig.~\protect\ref{fig17},
show a much wider $p$-distribution, which has intriguing consequences as described
below.

Next let us choose a fixed $k_0 '$-value, $k_0 '$ = 650 MeV/c,
and again fixed angles $ \theta_e $ = 40~$^\circ$, $ \phi_e $ = 0~$^\circ$,
$ \theta_n $ = 49.48~$^\circ$,
and $ \phi_n $ = 180~$^\circ$
all chosen out of the large domain $\Omega$. In Fig.~\protect\ref{fig625oben}
  we display the magnitudes of one of the
 amplitudes $\Lambda_\perp^{T'}$, now as a function of $p_n$. The others are
qualitatively similar. We compare different approximate treatments of 
the final state to the full calculation. 
The pure PWIA drops strongly with decreasing $p_n$. This is a simple consequence of the fact 
that the $^3$He wave function drops with increasing momenta. 
Choosing Jacobi momenta as arguments of the $^3$He wave function the photon momentum 
$\vec Q$ enters as $ \vec q = \vec{p}_n - \vec Q$, where $\vec q$ is the relative momentum
of the neutron in relation to the two protons. The decrease of the $^3$He wave function 
with increasing $q$ explains the PWIA curve in Fig.~\protect\ref{fig625oben}.
In case of the symmetrized PWIAS the photon can also be absorbed by the two protons,
which leads to $\vec q = \vec{p}_n $ and the occurrence of $\vec Q$ in the other 
Jacobi momentum $ \vec{p}_{23} $ as $ \vec{p}_{23} = \vec{p} \mp \frac12 \vec Q$.
As a consequence the two additional amplitudes in PWIAS start to contribute at lower 
$p_n$-values, which can clearly be seen in Fig.~\protect\ref{fig625oben}.
The curve denoted as FSI23 is based on the final state interaction among the two final
protons. This reduced final state interaction has apparently a strong effect
even near the quasi elastic peak. Again including full antisymmetrisation in the final 
state, but keeping only a first order final state interaction, leads to strong deviations
at low neutron momenta. This is denoted by FSI23S 
in Figs.~\protect\ref{fig625oben}-\protect\ref{fig627unten}.
Finally the full FSI (including of course antisymmetrisation in the final 
state) leads to a behavior, which is similar to FSI23 near the upper end of $p_n$ 
but deviates then from all other curves for lower $p_n$ values. 

Since the experiment under discussion emphasizes the large $p_n$-values 
in accordance with at least approximate quasi-free scattering conditions,
 we display in Figs.~\protect\ref{fig625unten}-\protect\ref{fig627unten} 
the magnitudes of the four $\Lambda_\beta^{\alpha}$ amplitudes
restricted to the domain seen in the experiment. 
We see a coincidence of PWIA and PWIAS in the restricted $p_n$ interval
and a spread of curves for the other cases. Especially the FSI is clearly 
distinct from FSI23 for $ R_\parallel^{TL'} $ and  $ R_\perp^{TL'} $.

The ratios of asymmetries for point geometries inside the domain $\Omega$  vary 
very much and depend extremely strongly on the treatment of the final
 3N state. A few more or less arbitrarily chosen cases are displayed 
in Figs.~\protect\ref{asym1}-\protect\ref{asym3}. In each case we see the ratio
for PWIA, PWIAS, FSI23, FSI and an additional case, FSIn. In the latter case
we put the electric form factor of the proton, $G_E^p$, to zero (the contribution
of $G_M^p$ is insignificant~\cite{Ziemer}). This has been done to demonstrate
the presence and importance of the photon absorption on the two protons.
Consequently averaging over asymmetries related to point geometries
 is not advisable.
Instead summing up cross sections first as in Eq.~(\ref{eq:sigmaOmega}) 
and then forming asymmetries is what has to be done.

Let us now show our results for the ratio of asymmetries given 
in Eq.~(\ref{eq:V}).
We parametrize $G_E^n$ by multiplying three models for $G_E^n$ by a factor $\lambda$.
We choose the ones by Gari-Kr\"umpelmann~\cite{Gari}. 
Fig.~\protect\ref{figV} shows
various theoretical ratios $V$ against $\lambda$ in comparison to the
experimental value of $ V^{\rm exp}$ = ( -7.26 $\pm$ 1.14 ) \%~\cite{BeckerPhD,triplestar}.
The largest $\lambda$-value results for FSI, followed by PWIAS, then PWIA
and finally FSI23. The four results for 
$G_E^n \equiv \lambda \cdot G_E^n\vert_{\rm model}$ 
are
 plotted in Fig.~\protect\ref{fig712} in the range of $Q^2$ 
values touched in that experiment. This refers to one of the three models.
 Finally we show in Fig.~\protect\ref{fig710unten}
$G_E^n$ as extracted through FSI and including the spread caused by the
experimental error. 
Superimposed on the spread caused by the experimental error 
we see small variations due to the three different choices of $G_E^n$-models.
For the central values around $Q^2$ = 0.40 ${\rm GeV}^2/c^2$
that model dependence is totally
negligible. 
All our results are displayed in Table~\protect\ref{table3},
where we have taken the average of the highest and lowest values 
in Fig.~\protect\ref{fig710unten}.
This average thus takes into account uncertainties of both,
model dependence for $G_E^n$ and experimental errors.

It is astonishing that PWIAS (and PWIA) are relatively close to the 
value based on FSI. This is due to an accidental conspiracy. The
slower decrease in the $p$-distribution for PWIA(S) shown 
in Fig.~\protect\ref{fig17}
causes smaller energies of the scattered electrons than for FSI.
As a consequence the photon-direction deviates
more strongly in case of $ \vec{S}_\perp $ from 90~$^\circ$ than for FSI. 
This leads
to a strongly modified contribution $ ( \int d \Omega v_{T'} R^{T'})_\perp $
for PWIAS in comparison to using FSI. 
In addition because of the lacking FSI
there are smaller protonic contributions. This together, as 
a detailed investigation shows~\cite{Ziemer}, 
yields the accidental result, that
 PWIAS is close to the full result.
The fact that PWIA(S) yields an unrealistic result could be verified by measuring 
the $p$-distribution for the response functions.

Our final result for FSI, $G_E^n$ = 0.052 $\pm$ 0.0038 at $Q^2$ = 0.40 ${\rm GeV}^2/c^2$,
is added in Fig.~\protect\ref{final} to the ones extracted 
from processes on the deuteron. There is a fair agreement.
Also added is another result achieved 
at MAMI~\cite{Rohe} at a higher $Q^2$-value. No FSI corrections have 
been taken into account in that case.

Our present result leaves room for improvement. The effect of MEC's 
like in section II is still to  be
explored and due to the relatively high $Q^2$-value one cannot exclude 
that relativistic effects might be noticeable. This is left to future 
investigations.

\section{Summary and Outlook}
\label{secIV}

We have extracted from two measurements on
$\overrightarrow{^3{\rm He}}(\vec{e},e')$ 
and 
$\overrightarrow{^3{\rm He}}(\vec{e},e'n)$ 
the magnetic and electric neutron form factors at certain $Q^2$-values.
 Our results are based on consistent Faddeev solutions for the 3N continuum 
and the 3N bound state. Modern NN forces have been used. In case of
 the inclusive reaction leading to $G_M^n$ we added $\pi$- and $\rho$-like 
two-body
 exchange currents
to the single nucleon current. Their effects were substantial. In the
exclusive process only a single
nucleon current operator has been used, which leaves room for improvement.
 In both
cases a strictly nonrelativistic formulation has been used, which also
 should be
improved. Our values for $G_M^n$ and $G_E^n$ agree well with the values
 extracted from processes on the deuteron.

From 3N scattering it is known that the most modern data-equivalent
 NN forces lead in nearly
all cases to results, which are very close together. We consider this
 robustness to be an important insight which gives
confidence to those choices of the 3N Hamiltonian. In the case of
 photon-induced processes a corresponding verification of robustness
 against interchanges of NN forces
and consistent MEC's is still missing. This refers not only to the 3N
 system but to the 2N system as well. Also generally accepted and
 feasible relativistic
formalisms have still to be worked out.

\acknowledgements

This work was supported by
the Deutsche Forschungsgemeinschaft (J.G. and H.K.),
the Polish Committee for Scientific Research under Grant No. 2P03B03914
and the Science and Technology Cooperation Germany-Poland. 
One of us (W.G.) would like to thank the Foundation for Polish Science
for the financial support during his stay in Cracow. We would like to thank 
Dr. H. Gao and Dr. D. Dutta for providing us with detailed information
about their experiment and data analysis.
The numerical calculations have been performed
on the PVP machines of the NERSC, USA, 
on the Cray T90 of the NIC in J\"ulich, Germany,
and on the VPP300 machine of the RWTH in Aachen, Germany.


\begin{table}[hbt]
\begin{center}
\begin{tabular}{|c|c|c|} \hline
 $Q^{2}$ (GeV/c)$^{2}$ & $G_{M}^{n}/G_{M}^{n}(Dipole)$ & Uncertainties \\
\hline
 0.1    &  0.966    &$\pm$0.014$\pm$0.01 \\
 0.2    &  0.962    &$\pm$0.013$\pm$0.01 \\
\hline
\end{tabular}
\end{center}
\caption[]{$G_{M}^{n}$ as a function of $Q^{2}$, the uncertainties are
statistical and systematic.}
\label{gmn_tbl}
\end{table}

\begin{table}[hbt]
\begin{center}
\begin{tabular}{|c|c|c|}
\hline
 $\theta_e$ $[{^\circ}]$ & $p_n^{min}$ [MeV/c] & $p_n^{max}$ [MeV/c] \\
\hline
    39 & 500.31 & 598.68 \\
    40 & 507.60 & 609.08 \\
    41 & 515.14 & 619.88 \\
    42 & 522.93 & 631.15 \\
    43 & 530.99 & 642.90 \\
    44 & 539.33 & 655.18 \\
    45 & 547.98 & 668.00 \\
    46 & 556.94 & 681.43 \\
    47 & 566.24 & 695.52 \\
    48 & 575.91 & 704.73 \\
    49 & 585.95 & 714.23 \\
    50 & 596.40 & 724.04 \\
    51 & 607.28 & 734.18 \\
    52 & 618.65 & 744.66 \\
    53 & 630.51 & 755.50 \\
    54 & 642.90 & 766.74 \\
    55 & 651.02 & 772.61 \\
    56 & 659.38 & 778.61 \\
    57 & 668.00 & 784.69 \\
    58 & 676.90 & 790.91 \\
    59 & 686.06 & 797.24 \\ 
\hline
\end{tabular}
\end{center}
\caption[]
{Intervals for the neutron momenta.}
\label{tab71}
\end{table}

\begin{table}[hbt]
\begin{center}
\begin{tabular}{|l|ccc|} \hline

             &  \multicolumn{3}{c|}{\rule[-3mm]{0mm}{10mm}
                 $Q^2 \ \ [ \left( GeV/c \right) ^2 ] $}   \\
             &         0.30         &         0.35          &        0.40         \\ \hline
PWIA         & 0.0441 $\pm$ 0.0035  & 0.0465 $\pm$ 0.0038  & 0.0484 $\pm$ 0.0038  \\
PWIAS        & 0.0455 $\pm$ 0.0035  & 0.0480 $\pm$ 0.0038  & 0.0499 $\pm$ 0.0038  \\
FSI23        & 0.0406 $\pm$ 0.0035  & 0.0428 $\pm$ 0.0037  & 0.0446 $\pm$ 0.0038  \\
FSI          & 0.0474 $\pm$ 0.0036  & 0.0499 $\pm$ 0.0038  & 0.0520 $\pm$ 0.0038  \\ \hline \hline

             &  \multicolumn{3}{c|}{\rule[-3mm]{0mm}{10mm}
                 $Q^2 \ \ [ \left( GeV/c \right) ^2 ] $}   \\
             &         0.45         &         0.50          &        0.55          \\ \hline
PWIA         & 0.0500 $\pm$ 0.0039  & 0.0512 $\pm$ 0.0043  & 0.0523 $\pm$ 0.0049  \\
PWIAS        & 0.0515 $\pm$ 0.0038  & 0.0529 $\pm$ 0.0043  & 0.0539 $\pm$ 0.0049  \\
FSI23        & 0.0460 $\pm$ 0.0038  & 0.0472 $\pm$ 0.0043  & 0.0482 $\pm$ 0.0048  \\
FSI          & 0.0536 $\pm$ 0.0039  & 0.0550 $\pm$ 0.0044  & 0.0561 $\pm$ 0.0050  \\ \hline
\end{tabular}
\end{center}
\caption[]
{Extracted averaged $G_E^n$ values (see text) obtained for different assumptions
about the final state. The uncertainties arise from the spread 
in the form factor parametrization and the experimental error.} 
\label{table3}
\end{table}


\begin{figure}[h!]
\epsfbox{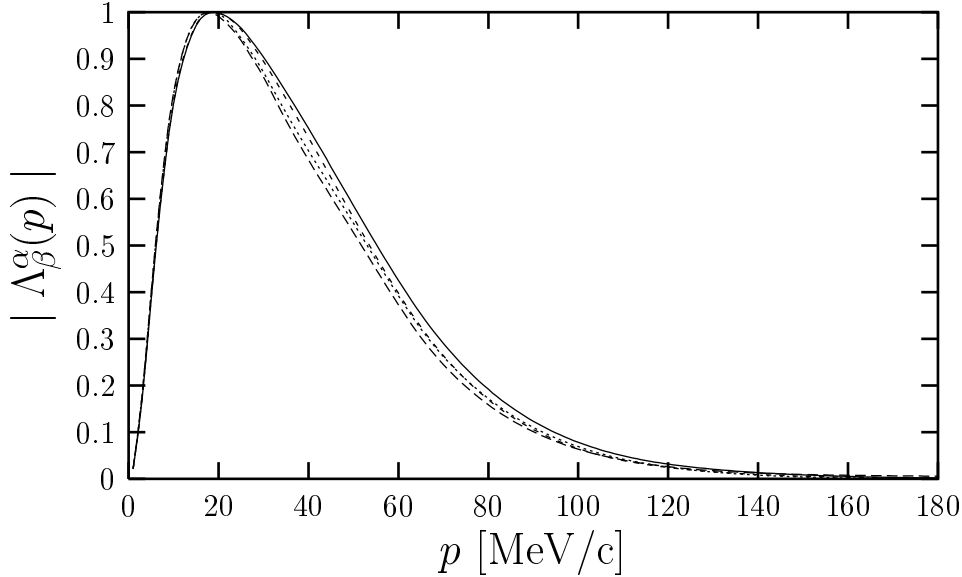}
\caption[ ]{The magnitudes of the $ \Lambda_\parallel^{T'} (p) $ (solid), 
$ \Lambda_\parallel^{TL'} (p) $ (long dashed),
$ \Lambda_\perp^{T'} (p) $ (short dashed) and
$ \Lambda_\perp^{TL'} (p) $ (dotted) amplitudes
of Eq.~(\protect\ref{eq:Lambda})
for full FSI as a function of $p$.
They are all arbitrarily normalised to 1 at their maxima and correspond
to the arbitrarily chosen values of 
$\theta_e$ = 43 $^\circ$, $\phi_e$ = 0 $^\circ$,
$\theta_n$ = 53.9 $^\circ$, $\phi_n$ = 180 $^\circ$ and $p_n$ = 530 MeV/c.}
\label{fig16}
\end{figure}

\begin{figure}[h!]
\epsfbox{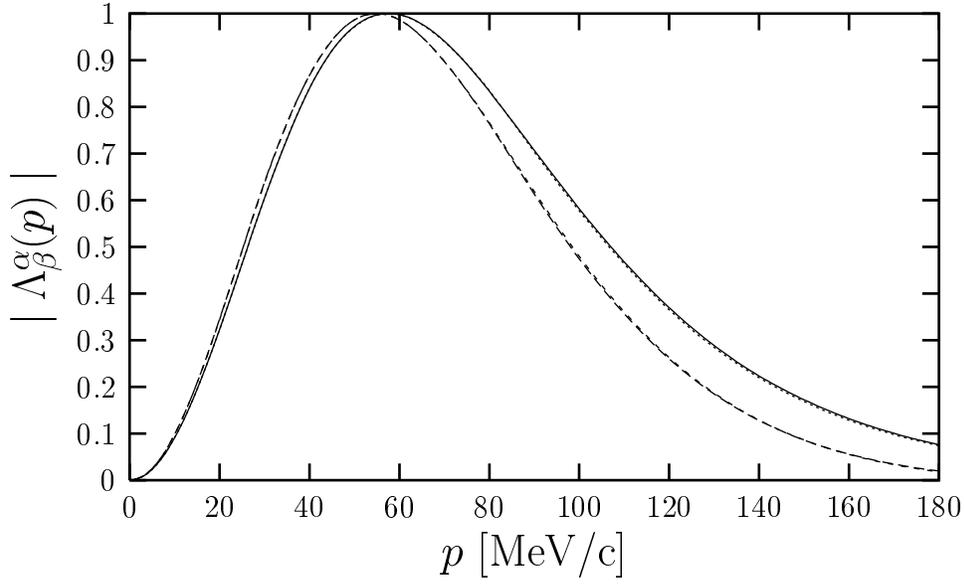}
\caption[ ]{The same as in Fig.~\protect\ref{fig16} for PWIA.}
\label{fig17}
\end{figure}

\begin{figure}[h!]
\epsfbox{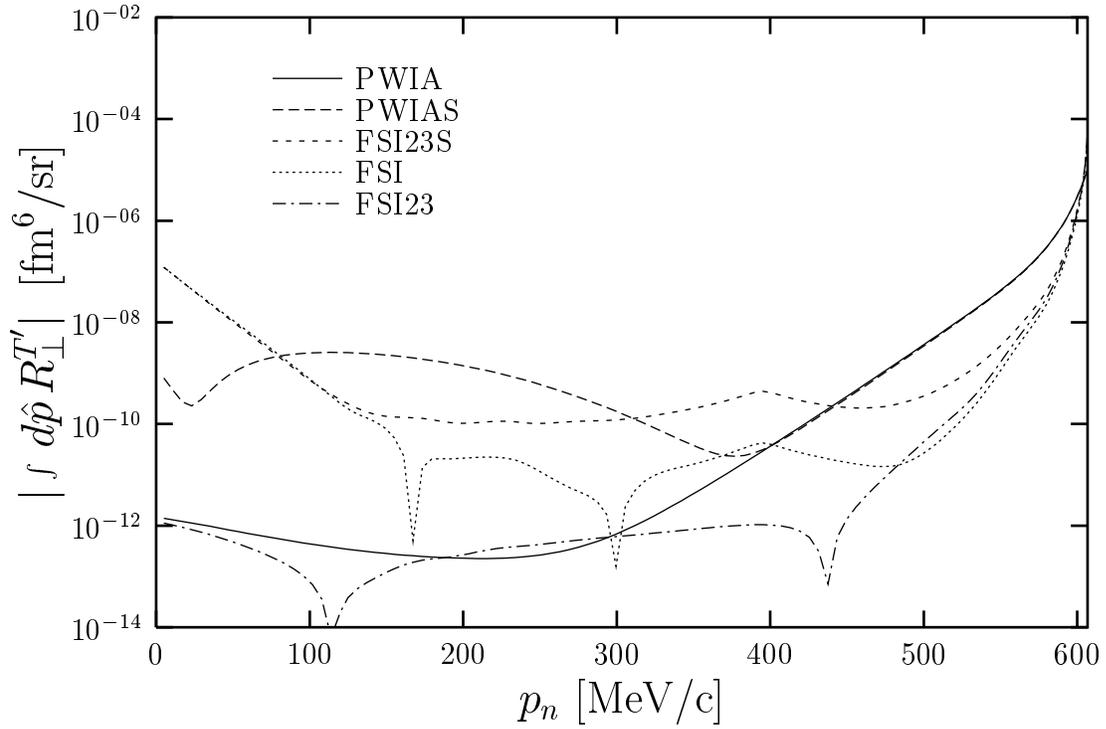}
\caption[ ]{FSI effects in the integrated response function $R^{T'}$.}
\label{fig625oben}
\end{figure}

\begin{figure}[h!]
\epsfbox{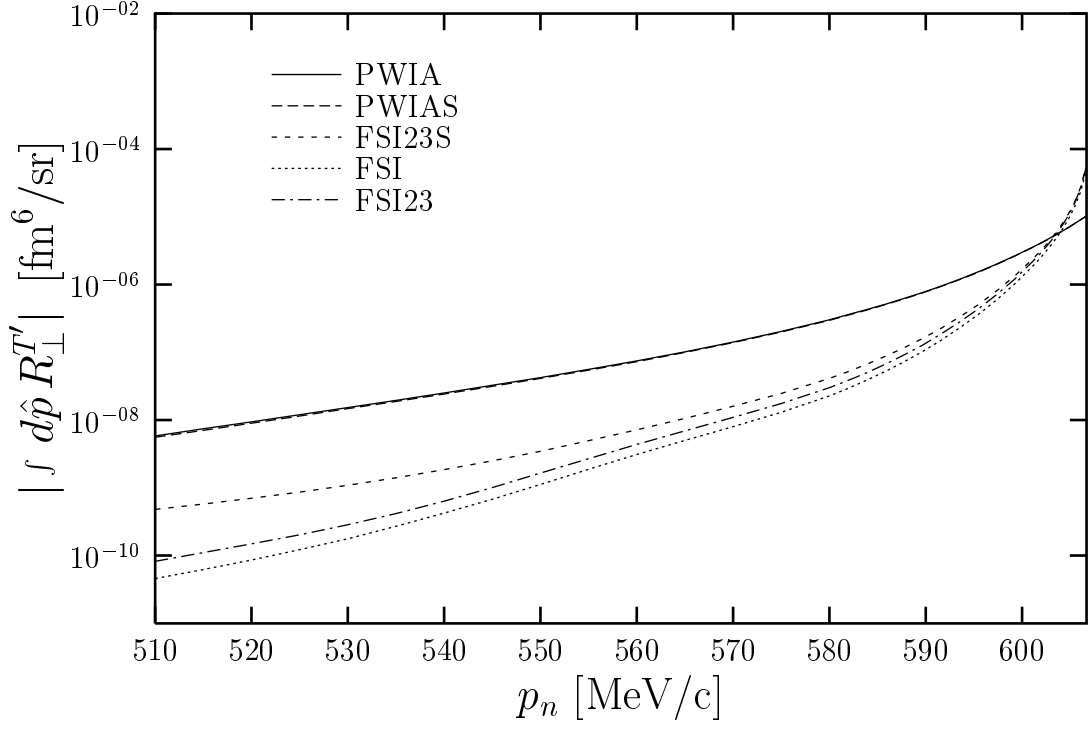}
\caption[ ]{The same as in Fig.~\protect\ref{fig625oben} for the truncated 
$p_n$ region used in the analysis of the experiment~\protect\cite{triplestar}.}
\label{fig625unten}
\end{figure}

\begin{figure}[h!]
\epsfbox{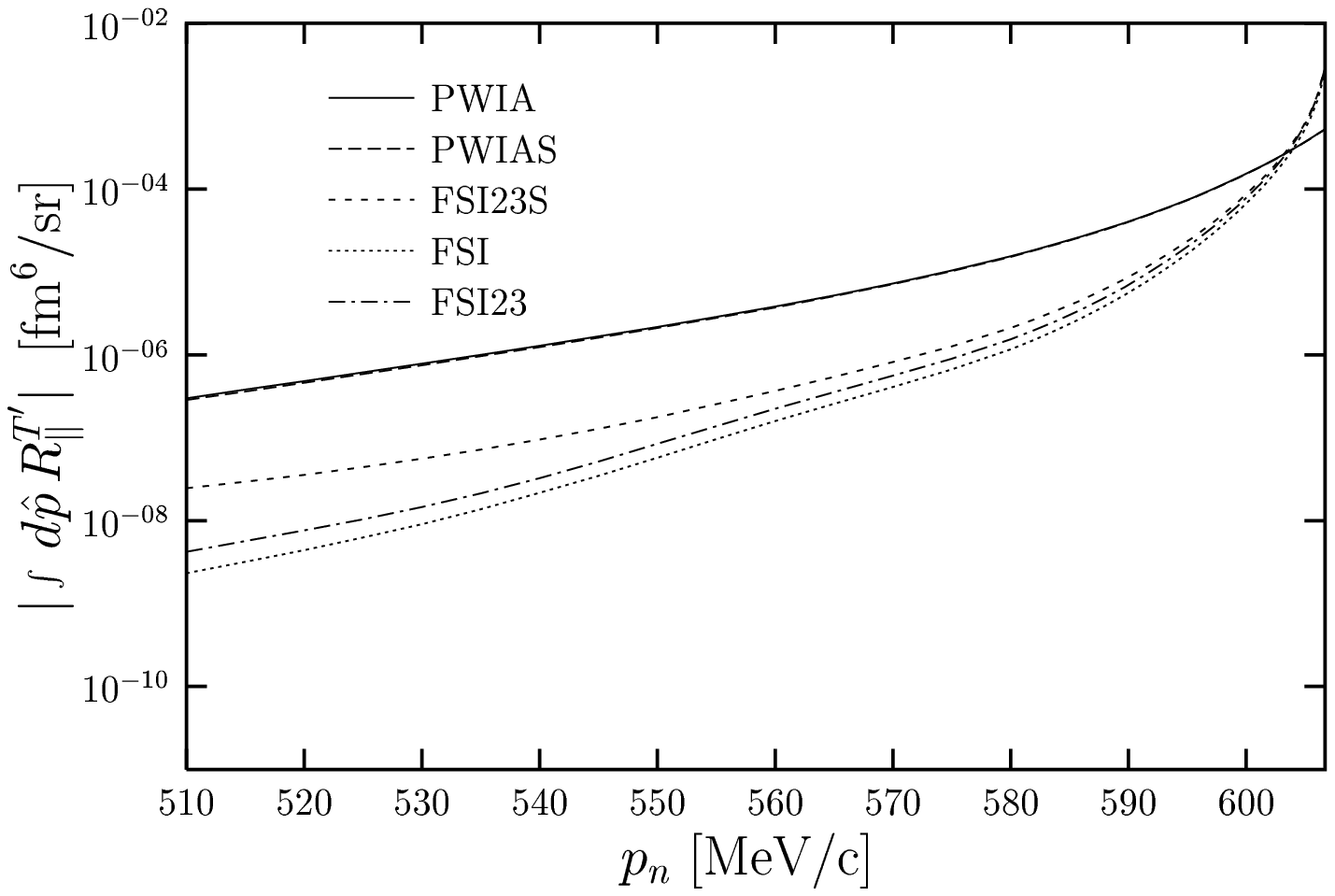}
\caption[ ]{The same as in Fig.~\protect\ref{fig625unten} but for the parallel
orientation of the target spin.}
\label{fig624unten}
\end{figure}

\begin{figure}[h!]
\epsfbox{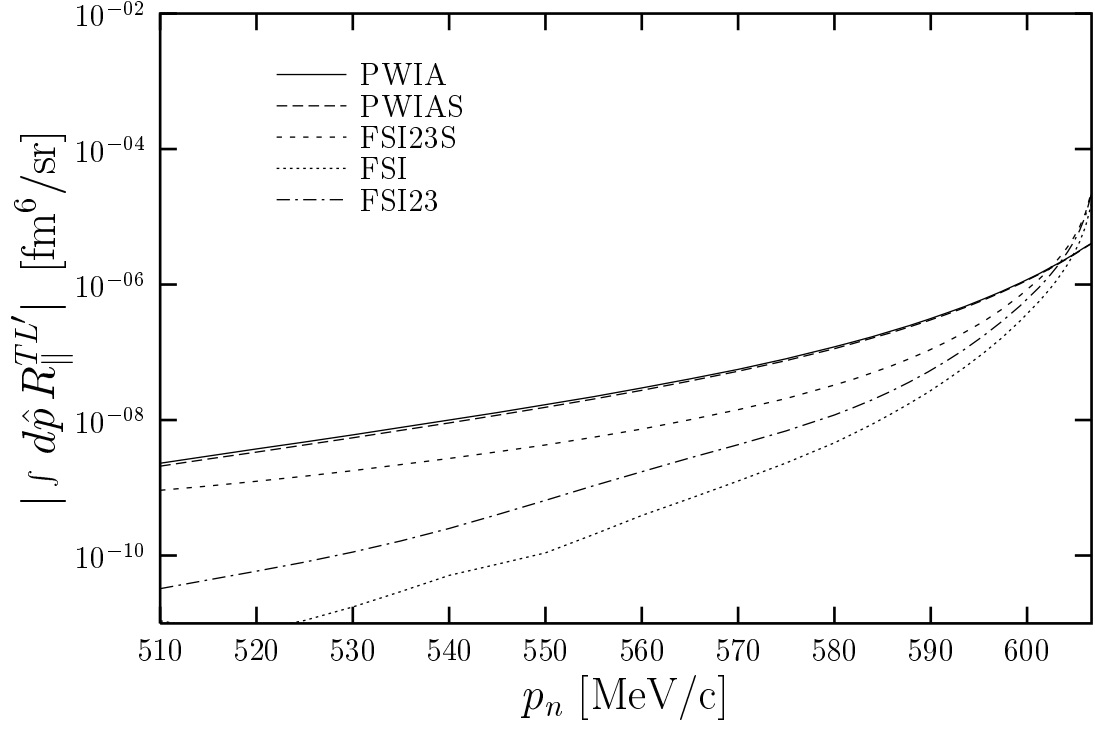}
\caption[ ]{The same as in Fig.~\protect\ref{fig624unten} 
but for the response function $R^{TL'}$.}
\label{fig626unten}
\end{figure}

\begin{figure}[h!]
\epsfbox{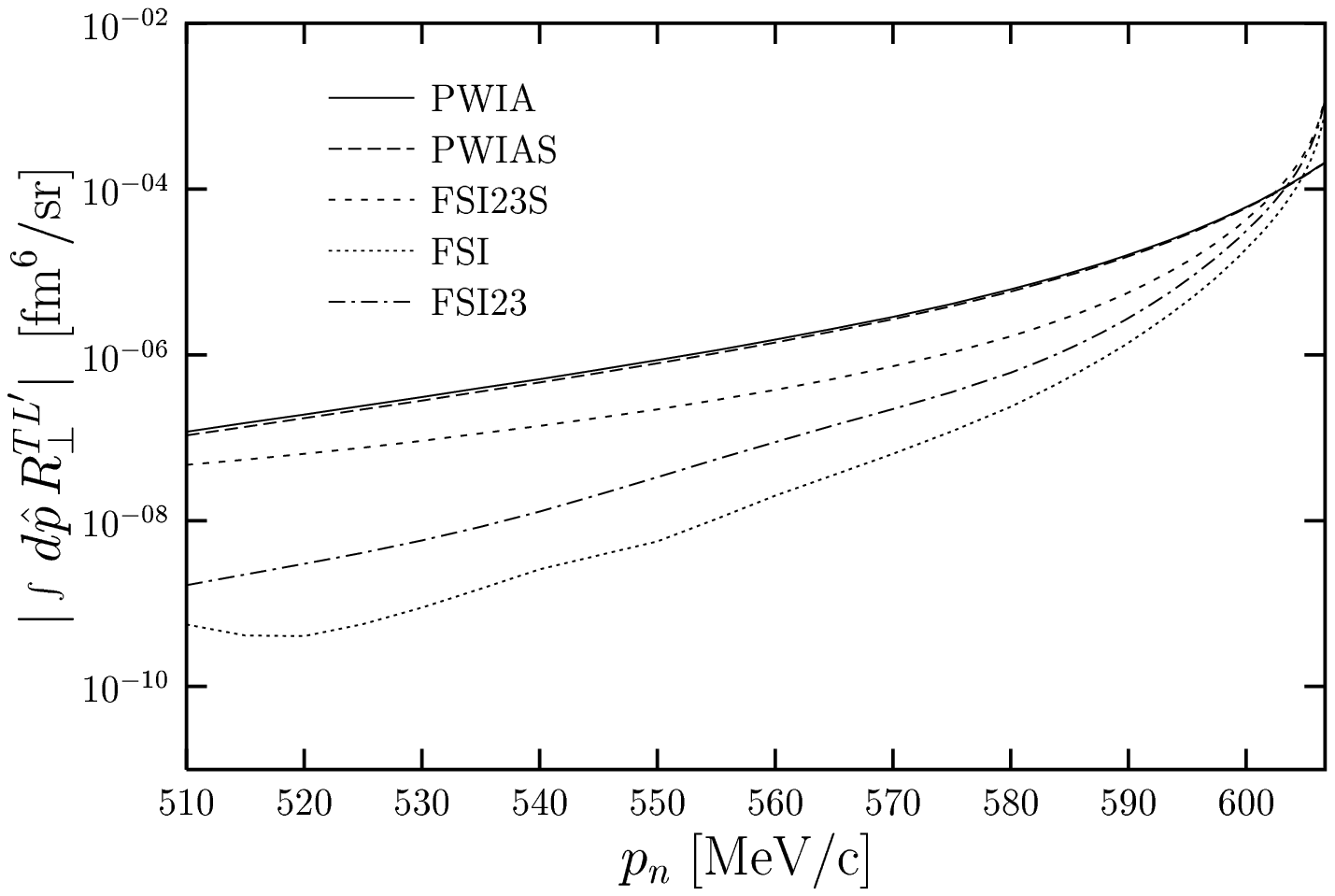}
\caption[ ]{The same as in Fig.~\protect\ref{fig626unten} but for the parallel
orientation of the target spin.}
\label{fig627unten}
\end{figure}

\begin{figure}[h!]
\epsfbox{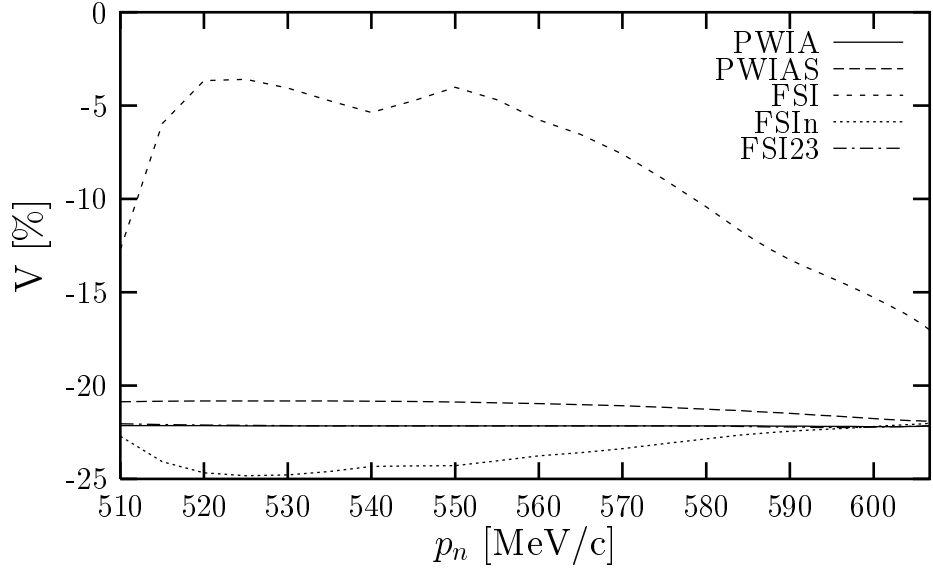}
\caption[ ]{Ratios $V$ for point geometry 
($k_0'$= 650 MeV/c, $\theta_e$= 40$^\circ$, $\theta_n$= 49.48$^\circ$, 
$\theta_\parallel^\star$= 1.12$^\circ$, $\theta_\perp^\star$= 88.88$^\circ$,
$\mid \vec Q \mid$= 549.61 MeV/c) for various treatments of the final state against $p_n$.}
\label{asym1}
\end{figure}

\begin{figure}[h!]
\epsfbox{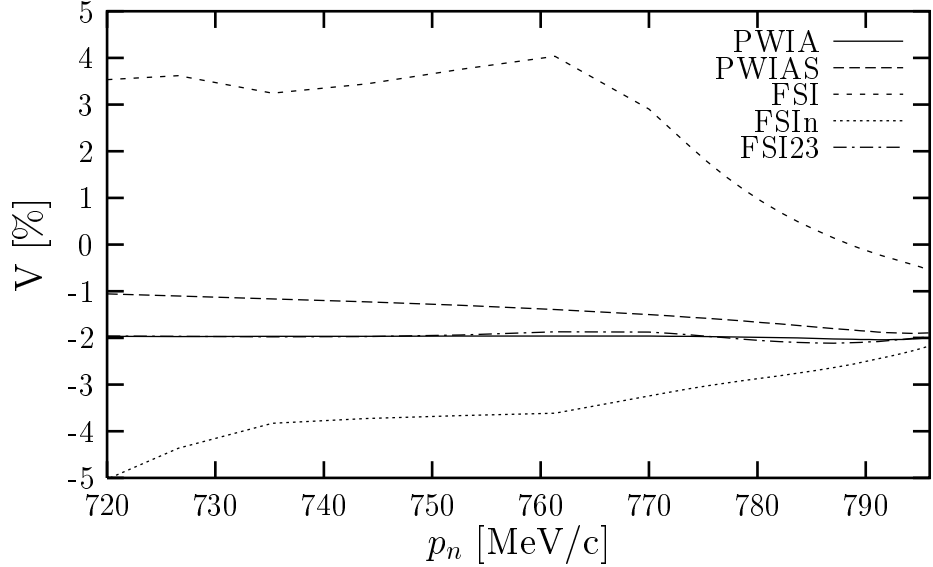}
\caption[ ]{The same as in Fig.~\protect\ref{asym1} for 
$k_0'$= 508 MeV/c, $\theta_e$= 58$^\circ$, $\theta_n$= 36.33$^\circ$, 
$\theta_\parallel^\star$= 14.27$^\circ$, $\theta_\perp^\star$= 75.73$^\circ$,
$\mid \vec Q \mid$= 727.16 MeV/c.}
\label{asym2}
\end{figure}

\begin{figure}[h!]
\epsfbox{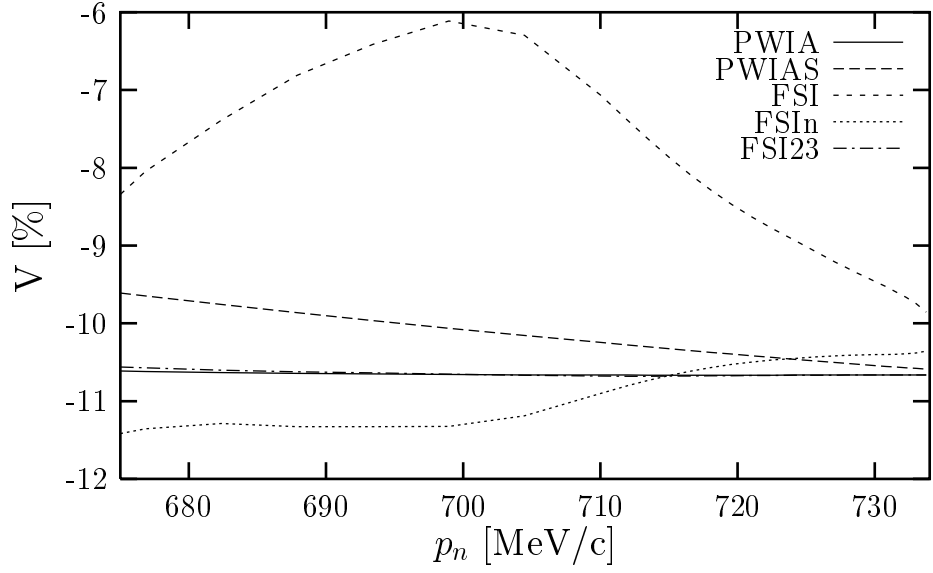}
\caption[ ]{The same as in Fig.~\protect\ref{asym1} for
$k_0'$= 560 MeV/c, $\theta_e$= 58$^\circ$, $\theta_n$= 40.93$^\circ$,     
$\theta_\parallel^\star$= 10.21$^\circ$, $\theta_\perp^\star$= 79.79$^\circ$,
$\mid \vec Q \mid$= 732.92 MeV/c.}
\label{asym3}
\end{figure}

\begin{figure}[h!]
\epsfbox{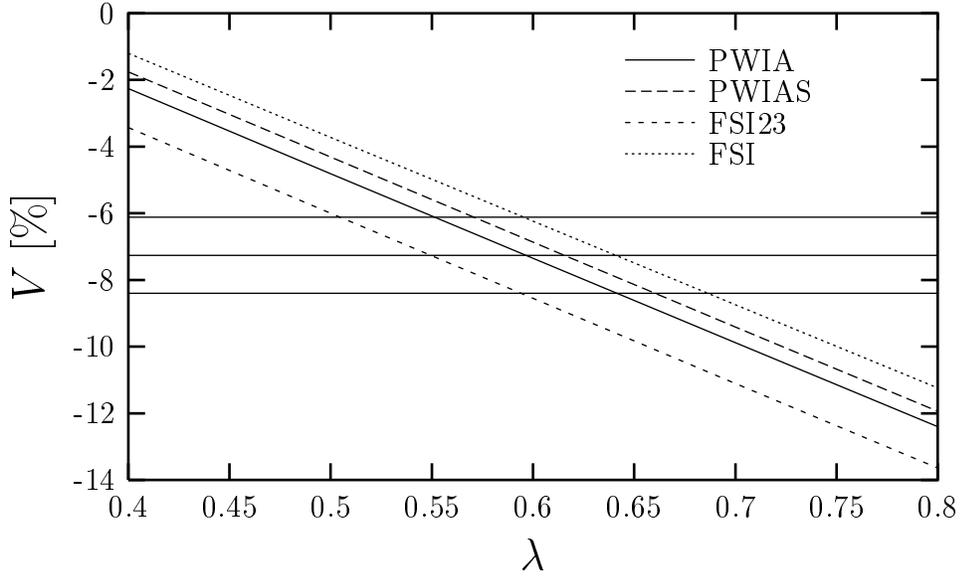}
\caption[ ]{Theoretical ratios $V$ against $\lambda$ in comparison to the experimental
value including its error.}
\label{figV}
\end{figure}

\begin{figure}[h!]
\epsfbox{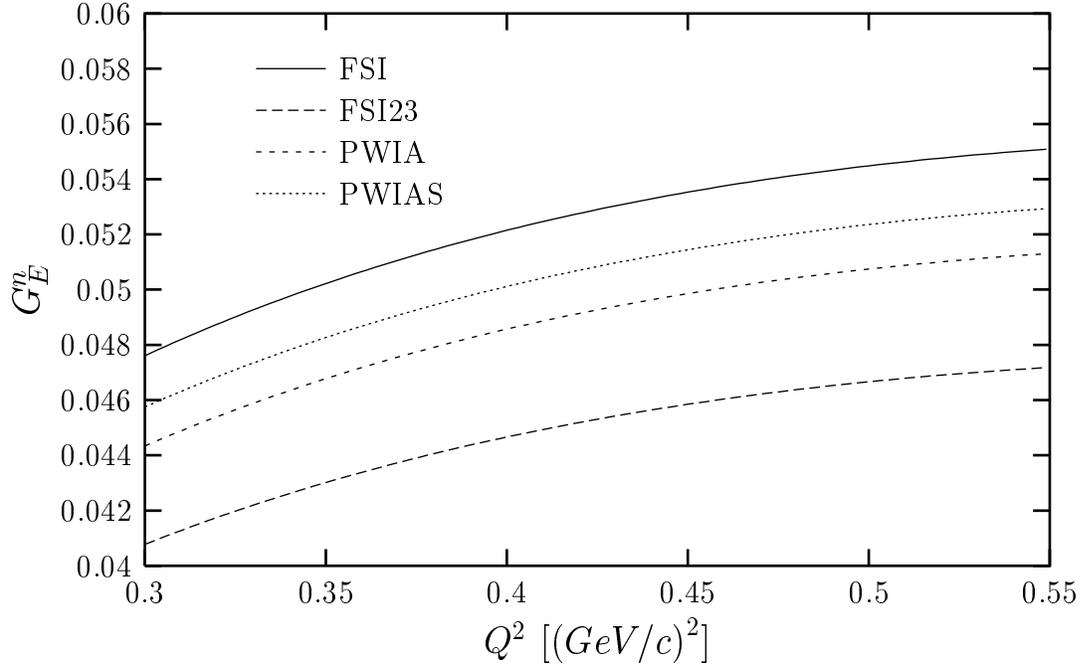}
\caption[ ]{Extracted $G_E^n$-values for different treatment of the final state.}
\label{fig712}
\end{figure}

\begin{figure}[h!]
\epsfbox{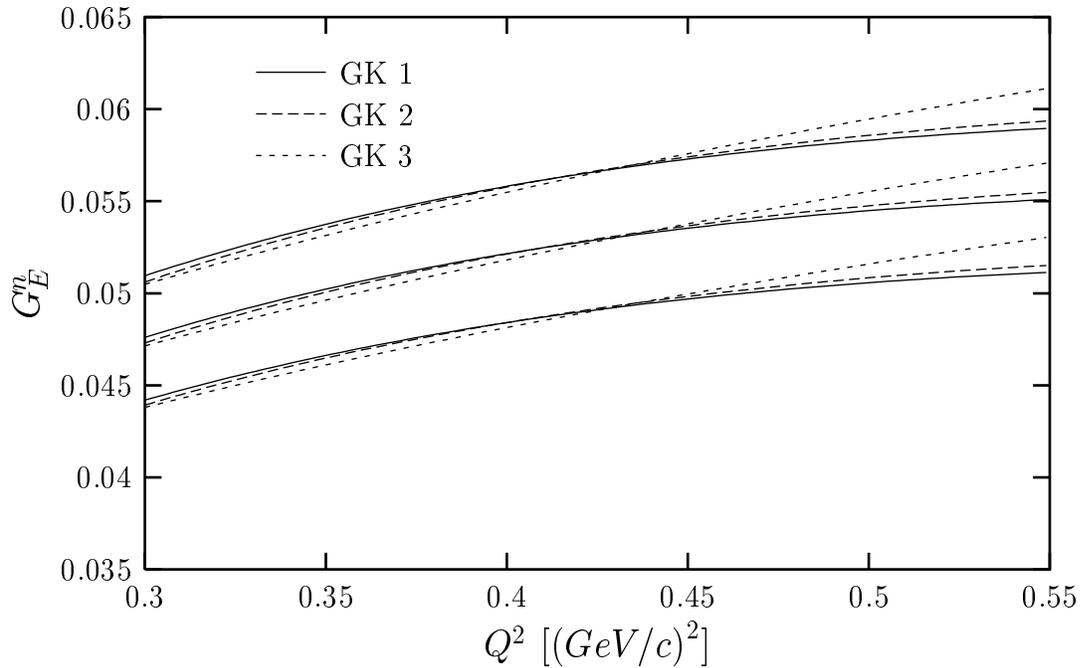}
\caption[ ]{The extracted $G_E^n$ for full FSI. The three separated curves 
correspond to the three $\lambda$-values from Fig.~\protect\ref{figV}.
The smaller spread of curves is due to the three different parametrisations 
of $\left. G_E^n\right|_{\rm model}$.}
\label{fig710unten}
\end{figure}

\begin{figure}[h!]
\epsfbox{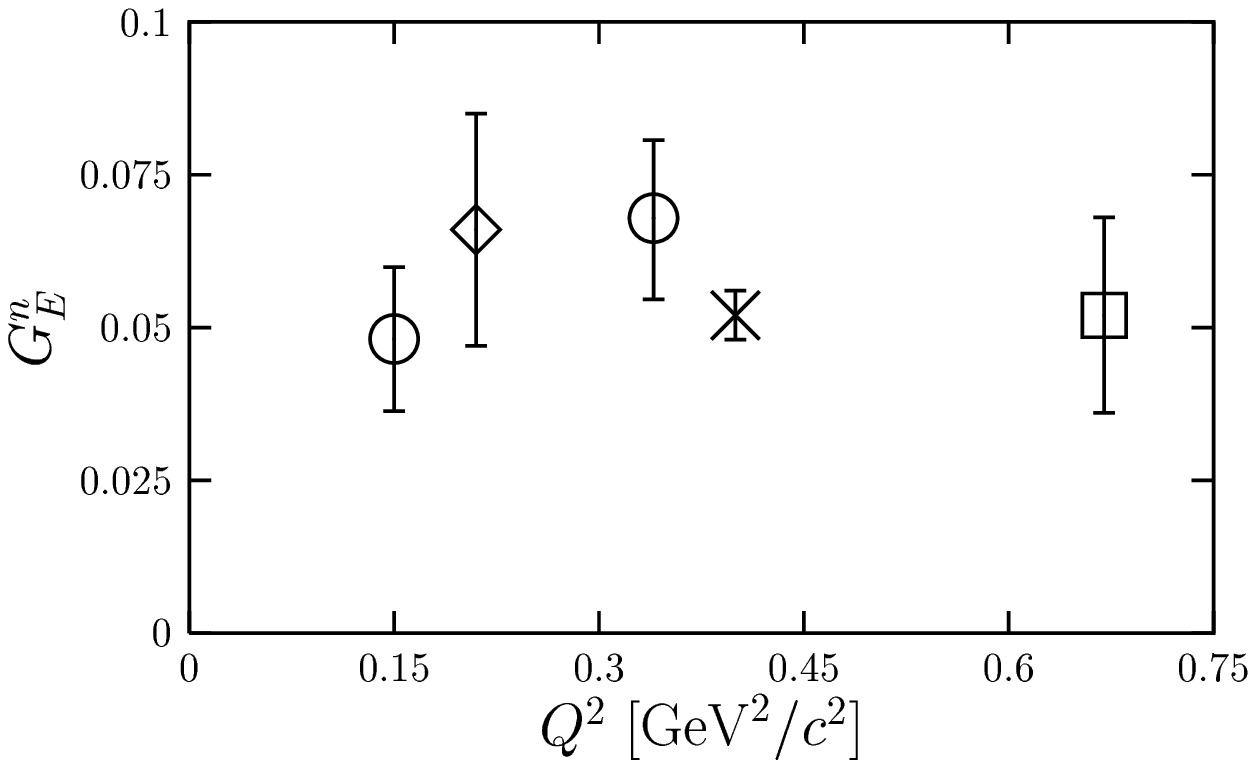}
\caption[ ]{$G_E^n$-values extracted from $^3$He 
(this work ($\times$), \protect\cite{Rohe} ($\Box$))
and from processes on the deuteron 
(\protect\cite{Passchier} ($\Diamond$), 
\protect\cite{Herberg} ($\bigcirc$)).}
\label{final}
\end{figure}

\end{document}